\documentclass[prb,reprint,aps,amsmath,amssymb,twocolumn,superscriptaddress,longbibliography]{revtex4-2}
\usepackage{times}
\usepackage{graphicx,bm,color,appendix}
\usepackage{float}
\definecolor{myyellow}{RGB}{255, 145, 0}
\usepackage[colorlinks,bookmarks=true,citecolor=blue,linkcolor=blue,urlcolor=blue]{hyperref}
\usepackage{amsmath,amssymb}
\usepackage{comment}
\usepackage{xcolor}
\usepackage[version=4]{mhchem}
\DeclareUnicodeCharacter{2212}{-}

\begin{document}
	
	\title{Electronically Inactive Intercalated \ce{La2NiO4} Layer in Superconducting \ce{La5Ni3O11}}
	\author{Tianyang Xie}
    \thanks{These authors contributed equally.}
    \affiliation{Beijing National Laboratory for Condensed Matter Physics and Institute of Physics, Chinese Academy of Sciences, Beijing 100190, China}
	
	\author{Yuxin Wang}
    \thanks{These authors contributed equally.}
	\affiliation{Kavli Institute for Theoretical Sciences, University of Chinese Academy of Sciences,
		Beijing, 100190, China}
	
	\author{Zhan Wang}
	\affiliation{Beijing National Laboratory for Condensed Matter Physics and Institute of Physics, Chinese Academy of Sciences, Beijing 100190, China}
	
	\author{Kun Jiang}
	\email{jiangkun@iphy.ac.cn}
	\affiliation{Beijing National Laboratory for Condensed Matter Physics and Institute of Physics, Chinese Academy of Sciences, Beijing 100190, China}
	\affiliation{School of Physical Sciences, University of Chinese Academy of Sciences, Beijing 100190, China}
	
	\author{Jiangping Hu}
	\email{jphu@iphy.ac.cn}
	\affiliation{New Cornerstone Science Laboratory, Beijing National Laboratory for Condensed Matter Physics and Institute of Physics,
		Chinese Academy of Sciences, Beijing 100190, China}
	
	\begin{abstract}
		The recent discovery of superconductivity in \ce{La5Ni3O11} extends the family of superconducting Ruddlesden--Popper nickelates beyond \ce{La3Ni2O7}. Unlike conventional members of a single Ruddlesden--Popper series, \ce{La5Ni3O11} contains an intercalated \ce{La2NiO4} layer between \ce{La3Ni2O7} blocks, raising the question of whether this additional layer participates in the low-energy electronic structure. Here, we combine density functional theory, Wannier-based tight-binding modeling, and rotationally invariant slave-boson calculations to investigate the electronic role of the intercalated layer. We find that realistic electronic parameters place the \ce{La2NiO4} layer in gapped insulating regimes rather than a paramagnetic metallic state. Furthermore, realistic interlayer hybridization fails to generate any appreciable \ce{La2NiO4}-derived spectral weight at the Fermi level. Our results demonstrate that the low-energy electronic structure of \ce{La5Ni3O11} is governed primarily by the \ce{La3Ni2O7} block, with the intercalated \ce{La2NiO4} layer remaining electronically inactive. This establishes a minimal low-energy description of \ce{La5Ni3O11} and provides a unified framework for understanding superconductivity in intercalated Ruddlesden--Popper nickelates.
	\end{abstract}
	\maketitle

		\section{Introduction} 
		
\begin{figure*}[htbp]
		\centering
		\includegraphics[width=0.9\textwidth]{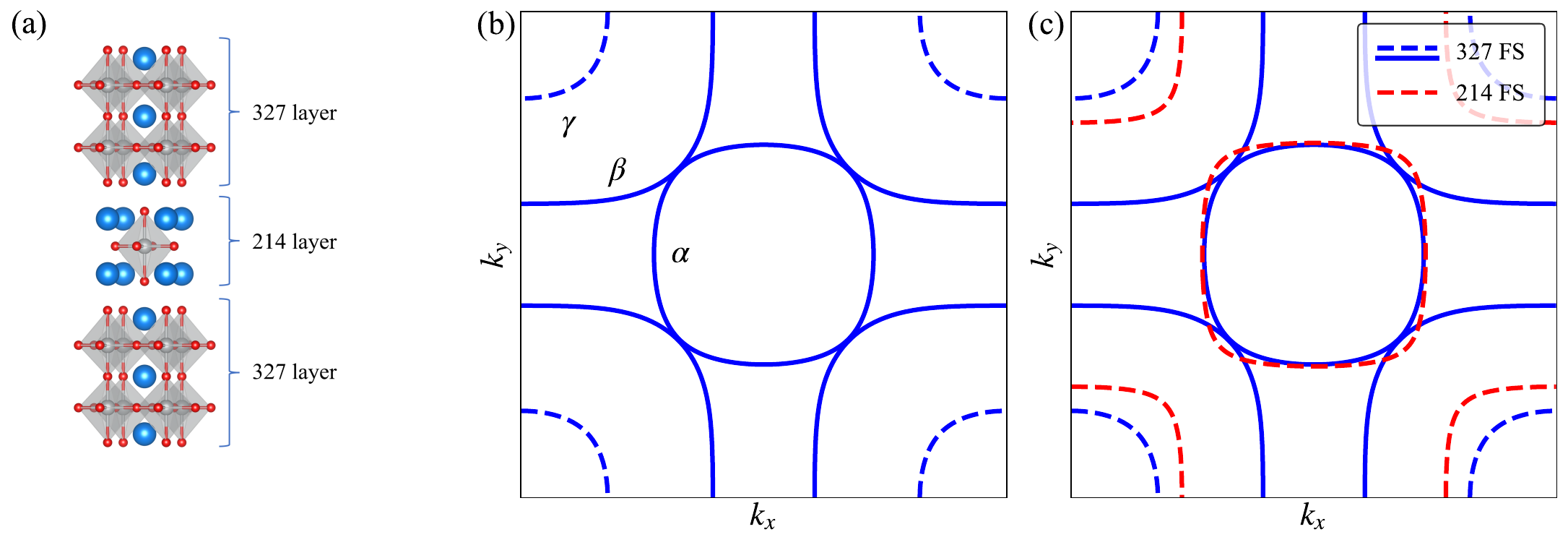} 
		\caption{(a) Crystal structure of \ce{La5Ni3O11}, consisting of alternating \ce{La3Ni2O7} (327) and intercalated \ce{La2NiO4} (214) blocks. The central question addressed in this work is whether the intercalated 214 layer contributes to the low-energy electronic structure of \ce{La5Ni3O11}. (b) Schematic Fermi surface of high-pressure \ce{La3Ni2O7}, consisting of one electron pocket ($\alpha$) and two hole pockets ($\beta$ and $\gamma$). The existence of the $\gamma$ pocket remains under debate ~\cite{327_DFT_01,327_ARPES_01,327_FLEX_01}. (c) Schematic Fermi surface of \ce{La5Ni3O11} if the intercalated 214 layer were electronically active ~\cite{5311_sc_01,5311_theory_RPA_01}. Compared with (b), two additional Fermi pockets originating from the 214 block (red) appear alongside the 327-derived pockets (blue). This work demonstrates that these additional 214-derived Fermi pockets are absent in the realistic low-energy electronic structure.}
        %(a) Crystal structure of \ce{La5Ni3O11}, consisting of alternating \ce{La2NiO4} and \ce{La3Ni2O7} layers. Electronically, whether \ce{La2NiO4} layer contributes to \ce{La5Ni3O11} is the central question we want to address. 
        %(b) Fermi surface of \ce{La3Ni2O7} at high-pressure. There are three Fermi pockets: one electron pocket $\alpha$ and two hole pockets $\gamma$ and $\beta$. Notice that the presence of a $\gamma$ Fermi surface remains under debate. (c) The proposed and DFT-calculated Fermi surfaces of \ce{La5Ni3O11}, where the blue (red) portions originate from the \ce{La3Ni2O7} (\ce{La2NiO4}) layers. There are two additional pockets from \ce{La2NiO4} block comparing to (b).}
		\label{fig:Fermi_surface}
	\end{figure*}

    The emergence of superconductivity in the Ruddlesden–Popper (RP) nickelate \ce{La3Ni2O7} under pressure has opened a new route toward high-temperature superconductivity. Following the initial discovery in \ce{La3Ni2O7} (327)~\cite{327_original_exp_01,327_original_exp_02,327_original_exp_03}, superconductivity has subsequently been observed in several related RP compounds, including \ce{La4Ni3O10} (4310)~\cite{4310_original_exp_01,4310_original_exp_02,4310_original_exp_03}, \ce{La5Ni3O11} (5311)~\cite{5311_sc_01}, as well as rare-earth-substituted variants~\cite{327_Pr_01,327_Sm_01}. Among these systems, \ce{La5Ni3O11} is particularly intriguing because it does not belong to a conventional single RP series. Instead, it represents a naturally hybrid structure composed of alternating \ce{La3Ni2O7} and \ce{La2NiO4} (214) building blocks~\cite{5311_structure_01}, as schematically illustrated in Fig.~\ref{fig:Fermi_surface}(a).
	
	This structural hybridity immediately raises a central question: what role does the intervening 214 layer play in the low-energy physics? Under pressure, \ce{La5Ni3O11} exhibits a phase diagram strikingly similar to that of bulk \ce{La3Ni2O7}, featuring density-wave behavior at low pressure and superconductivity at high pressure with a comparable transition temperature~\cite{5311_sc_01,327_original_exp_01,327_review_03,327_phase_diagram_01}. Transport measurements further reveal a pronounced anisotropy, with metallic in-plane conduction but insulating $c$-axis transport, pointing to an underlying quasi-two-dimensional electronic structure ~\cite{5311_sc_01}. Together, these observations suggest that the essential low-energy physics of \ce{La5Ni3O11} may largely originate from the 327 block. However, the embedded 214 layer could, in principle, introduce additional bands near the Fermi level and reconstruct the Fermi surface, as schematically illustrated in Fig.~\ref{fig:Fermi_surface}(b,c). 
    %Recent theoretical studies have reached conflicting conclusions regarding the role of the intercalated layer. Weak-coupling approaches predict that the intercalated single layer remains electronically active and contributes additional Fermi-surface states, whereas strong-coupling studies instead find pronounced layer selectivity, with the low-energy electronic structure dominated by the bilayer block ~\cite{5311_theory_RPA_01,5311_theory_RPA_02,5311_theory_DMFT_01,5311_theory_DMFT_02,5311_theory_DMFT_03}. Determining whether the 214 layer is electronically active is therefore essential for establishing a minimal low-energy description of \ce{La5Ni3O11}. 
    %\textcolor{red}{Recent theoretical studies have proposed different pictures: DFT/RPA-based approaches retain single-layer-derived Fermi-surface states, whereas DFT+DMFT calculations emphasize strong layer selectivity and bilayer-dominated low-energy spectra~\cite{5311_theory_RPA_01,5311_theory_RPA_02,5311_theory_DMFT_01,5311_theory_DMFT_02,5311_theory_DMFT_03}.}
    Recent theoretical studies have emphasized different aspects of the hybrid electronic structure: weak-coupling approaches often retain single-layer-derived Fermi-surface states, whereas strong-coupling studies suggest that orbital-selective correlations strongly suppress the coherent low-energy contribution of the single layer~\cite{5311_theory_RPA_01,5311_theory_RPA_02,5311_theory_DMFT_01,5311_theory_DMFT_02,5311_theory_DMFT_03}. These results motivate a direct examination of whether the intercalated 214 layer itself can form a coherent low-energy electronic subsystem in \ce{La5Ni3O11}.

    Two natural scenarios can be envisioned for the electronic structure of \ce{La5Ni3O11}. If orbital splitting and electron correlations drive the intercalated 214 layer into a gapped state, its contribution to the low-energy electronic structure is strongly suppressed. In this case, the Fermi surface is governed primarily by the 327 block, as schematically illustrated in Fig.~\ref{fig:Fermi_surface}(b). Alternatively, if the 214 layer remains metallic, it contributes additional low-energy carriers and gives rise to extra Fermi pockets, leading to the more complex Fermi surface shown in Fig.~\ref{fig:Fermi_surface}(c). The first scenario is plausible because bulk \ce{La2NiO4} is an insulating antiferromagnet at ambient pressure~\cite{214_magnetic_01,214_magnetic_02,214_magnetic_03,214_magnetic_04,214_magnetic_05,214_magnetic_06}. However, the 214 layer in \ce{La5Ni3O11} resides in a distinct structural environment, and possible interlayer hybridization mediated by Ni $d_{z^2}$ orbitals could substantially modify its electronic structure. Whether the intercalated 214 layer remains electronically active at low energies therefore remains an open question.

	In this work, we show that the embedded 214 layer in \ce{La5Ni3O11} is effectively removed from the low-energy sector. We construct a two-orbital tight-binding model for the 214 layer and investigate its correlated ground states using the rotationally invariant slave-boson (RISB) method~\cite{RISB_01,RISB_02}. The resulting phase diagram demonstrates that realistic parameters extracted from both the generalized gradient approximation (GGA) in the Perdew-Burke-Ernzerhof (PBE) form \cite{perdew1996generalized} and the HSE06 hybrid functional, which mixes Hartree-Fock exchange with Kohn-Sham density functional theory \cite{becke1993density,krukau2006influence}, place the 214 layer in gapped phases rather than a paramagnetic metallic state. We further examine a minimal coupled 327+214 model and find that interlayer hybridization fails to restore appreciable 214-derived spectral weight at the Fermi level. Our results establish a simple electronic picture of \ce{La5Ni3O11}: its low-energy physics is governed primarily by the \ce{La3Ni2O7} block, while the intercalated \ce{La2NiO4} layer remains electronically inactive.

    \section{Model and Results}

    We begin by considering the intercalated \ce{La2NiO4} layer embedded within \ce{La5Ni3O11}, as illustrated in Fig.~\ref{fig:Fermi_surface}(a). Owing to its coupling with the neighboring 327 blocks, the 214 layer is structurally compressed relative to bulk \ce{La2NiO4} ~\cite{5311_sc_01,5311_structure_01}. As summarized in Table~\ref{tab:214_lattice_parameters_01}, both the in-plane lattice constants and the Ni--apical oxygen ($\mathrm{O}_t$) bond length are reduced. Such structural modifications can substantially alter the crystal-field splitting and hopping amplitudes, thereby changing the low-energy electronic structure of the embedded 214 layer.

    The electronic structure of bulk \ce{La2NiO4} is known to depend sensitively on the choice of exchange-correlation functional \cite{takegahara1989electronic,rivero2009description}. To ensure that our conclusions are robust against this uncertainty, we perform density functional calculations using the Vienna \emph{ab initio} simulation package (VASP) with the projector augmented-wave (PAW) method \cite{kresse1996efficient,kresse1999ultrasoft}. We consider both the PBE \cite{perdew1996generalized} and HSE06 functionals \cite{becke1993density,krukau2006influence}, and subsequently Wannierized \cite{mostofi2008wannier90,marzari2012maximally} to construct corresponding low-energy models for subsequent many-body analysis. Additional computational details are provided in Appendix D.

    \begin{table}[htbp]
    \centering
    \caption{Structural parameters of the \ce{La2NiO4} block in bulk \ce{La2NiO4} and in the \ce{La5Ni3O11} heterostructure. Here, $d_{\mathrm{Ni}-\mathrm{O}_t}$ denotes the distance between the Ni atom and the apical oxygen atom O$_t$ in the Ni-O octahedron.}
    \label{tab:214_lattice_parameters_01}
    \setlength{\tabcolsep}{8pt}
    \begin{tabular}{c c c c c}
        \hline
        \hline
          & $a$ (\AA) & $b$ (\AA) & $c$ (\AA) & $d_{\mathrm{Ni}-\mathrm{O}_t}$ (\AA) \\
        \hline
        214 bulk     & 3.92  & 3.92  & 12.52 & 4.482 \\
        214 in 5311  & 3.845 & 3.845 & ---   & 4.442 \\
        \hline
        \hline
    \end{tabular}
\end{table}

    %Hence, the tight-binding models for the intercalated 214-layer in \ce{La5Ni3O11} are constructed.	
    For the 214-layer in \ce{La5Ni3O11}, we retain the Ni $d_{x^2-y^2}$ and $d_{z^2}$ orbitals. 
    Then, the tight-binding (TB) Hamiltonian is constructed as
    \begin{align}
        H_0 =& \sum_{\alpha=x,z}\sum_{{\bm k}, \sigma=\uparrow,\downarrow} \Big[ \epsilon^\alpha + t_{1}^\alpha \gamma_{\bm k} + t_{2}^\alpha \gamma'_{\bm k}+ t_{3}^\alpha \gamma''_{\bm k} \Big] d_{{\bm k},\alpha,\sigma}^\dagger d_{{\bm k},\alpha,\sigma}\notag\\
        +& \sum_{{\bm k}, \sigma} \Big[ t_{1}^{xz} \beta_{\bm k}+ t_{3}^{xz}\beta'_{\bm k} \Big]
        \left(d_{{\bm k},x,\sigma}^\dagger d_{{\bm k},z,\sigma} + \mathrm{h.c.}\right),
	   \label{eq:214_Hamiltonian_01}
    \end{align}
    where $\gamma_{\bm k}=2(\cos k_x + \cos k_y)$, $\gamma'_{\bm k} = 4 \cos k_x \cos k_y$, $\gamma''_{\bm k}=2(\cos 2k_x + \cos 2k_y)$, $\beta_{\bm k}=2(\cos k_x-\cos k_y)$, and $\beta'_{\bm k}=2(\cos 2k_x - \cos 2k_y)$. Here, $d_{{\bm k}\alpha\sigma}^\dagger$ creates an electron in spin index $\sigma=\uparrow,\downarrow$, and orbital index $\alpha=x,z$, denoting $d_{x^2-y^2}$ and $d_{z^2}$, respectively. 

    The tight-binding parameters extracted from the density functional calculations are summarized in Table~\ref{tab:214_hopping_parameters}. Among them, the key parameter is the Jahn--Teller crystal-field splitting,
    \begin{equation}
        \Delta_{JT}=\epsilon^x-\epsilon^z,
    \end{equation}
    which governs the orbital occupation and the resulting low-energy electronic structure. While the hopping amplitudes obtained from the PBE and HSE06 functionals differ only by a few tens of meV, the corresponding values of $\Delta_{JT}$ differ by nearly 1~eV. This pronounced variation defines a realistic range of crystal-field splittings for the intercalated 214 layer.

    The comparison between the two functionals is particularly important because bulk \ce{La2NiO4} is a charge-transfer antiferromagnetic insulator whose electronic structure is highly sensitive to the treatment of exchange and correlation~\cite{214_photoemission_01,214_magnetic_01,214_magnetic_02,214_magnetic_03,214_magnetic_04,214_magnetic_05,214_magnetic_06}. Semilocal PBE tends to underestimate orbital polarization and insulating behavior, whereas the hybrid HSE06 functional favors a larger crystal-field splitting and stronger localization. Consequently, the PBE and HSE06 parametrizations span the physically relevant regime of $\Delta_{JT}$, allowing us to assess whether the electronic inactivity of the intercalated 214 layer is robust against the intrinsic uncertainty of the underlying electronic structure.

\begin{table*}[htbp]
    \centering
    \caption{Tight-binding parameters for the \ce{La2NiO4}-layer model obtained from DFT band structures using PBE and HSE06 functionals. $\epsilon^x$ and $\epsilon^z$ denote the onsite energies of the Ni $d_{x^2-y^2}$ and $d_{z^2}$ orbitals, respectively, and $\Delta_{JT}=\epsilon^x-\epsilon^z$ is the corresponding orbital splitting energy. All values are given in units of eV.}
    \label{tab:214_hopping_parameters}
    \setlength{\tabcolsep}{4pt}
    \begin{tabular}{c c c c c c c c c c c c}
        \hline
        \hline
        & $t_{1}^x$ & $t_{1}^z$ & $t_{2}^x$ & $t_{2}^z$ & $t_{3}^x$ & $t_{3}^z$ 
        & $t_{1}^{xz}$ & $t_{3}^{xz}$ & $\epsilon^x$ & $\epsilon^z$ & $\Delta_{JT}$ \\
        \hline
        PBE & -0.4264 & -0.0760 &  0.0740 & -0.0111 & -0.0489 & -0.0082 
            & -0.1802 & -0.0186 &  0.4984 & -0.0587 & 0.5571 \\
        HSE06 & -0.4650 & -0.0952 &  0.0852 & -0.0136 & -0.0421 & -0.0091 
            & -0.2251 & -0.0201 &  2.0461 & -0.4880 & 2.5341 \\
        \hline
        \hline
    \end{tabular}
\end{table*}

Furthermore, the electron correlation effects are included through the local two-orbital Hubbard interaction
\begin{equation}
    \label{eq:214_Hamiltonian_02}
    \begin{aligned}
        H_{\mathrm{int}} &= U \sum_{i,\alpha} n_{i,\alpha,\uparrow} n_{i,\alpha,\downarrow} + U^\prime \sum_{i} n_{i,x} n_{i,z} \\
        &- J_H \sum_{i} \left( n_{i,x,\uparrow} n_{i,z,\uparrow} + n_{i,x,\downarrow} n_{i,z,\downarrow} \right) \\
        &+ J_H \sum_{i,\sigma} d_{i,x,\sigma}^{\dagger} d_{i,z,\bar{\sigma}}^{\dagger} d_{i,x,\bar{\sigma}} d_{i,z,\sigma} \\
        &+ J_C \sum_{i} \left( d_{i,x,\uparrow}^{\dagger} d_{i,x,\downarrow}^{\dagger} d_{i,z,\downarrow} d_{i,z,\uparrow} + \mathrm{h.c.} \right),
    \end{aligned}
\end{equation}
where $U^\prime = U - 2J_H$ and $J_C = J_H$. 
Throughout this work, we adopt the commonly used parameter ratio $J_H/U = 0.1$ for multiorbital nickelate models~\cite{Hund_coupling_01,327_FLEX_01}. The Hamiltonian of the isolated 214 layer is then given by
\begin{equation}
    \label{eq:214_Hamiltonian_03}
    H^{214}=H_0+H_{\mathrm{int}}.
\end{equation}

To treat electronic correlations, we employ the multiorbital Kotliar--Ruckenstein (KR) slave-boson method ~\cite{KRSB_01}, also known as the rotationally invariant slave-boson formalism~\cite{RISB_01,RISB_02}. At the saddle-point level, KR slave boson is equivalent to the Gutzwiller approximation and captures both correlation-induced quasiparticle renormalization and symmetry-broken magnetic states. To determine the ground state of the isolated 214 layer, we consider three competing saddle-point solutions: a paramagnetic metal (PM), an antiferromagnetic (AFM) phase, and a band insulator (BI). For each solution, we analyze the total energy, electronic structure, charge gap, and staggered magnetic moment. Details of the RISB formalism and the treatment of the AFM phase are provided in Appendices~\ref{Sec:A1} and \ref{Sec:A2}.

\begin{figure}
    \centering
    \includegraphics[width=\columnwidth]{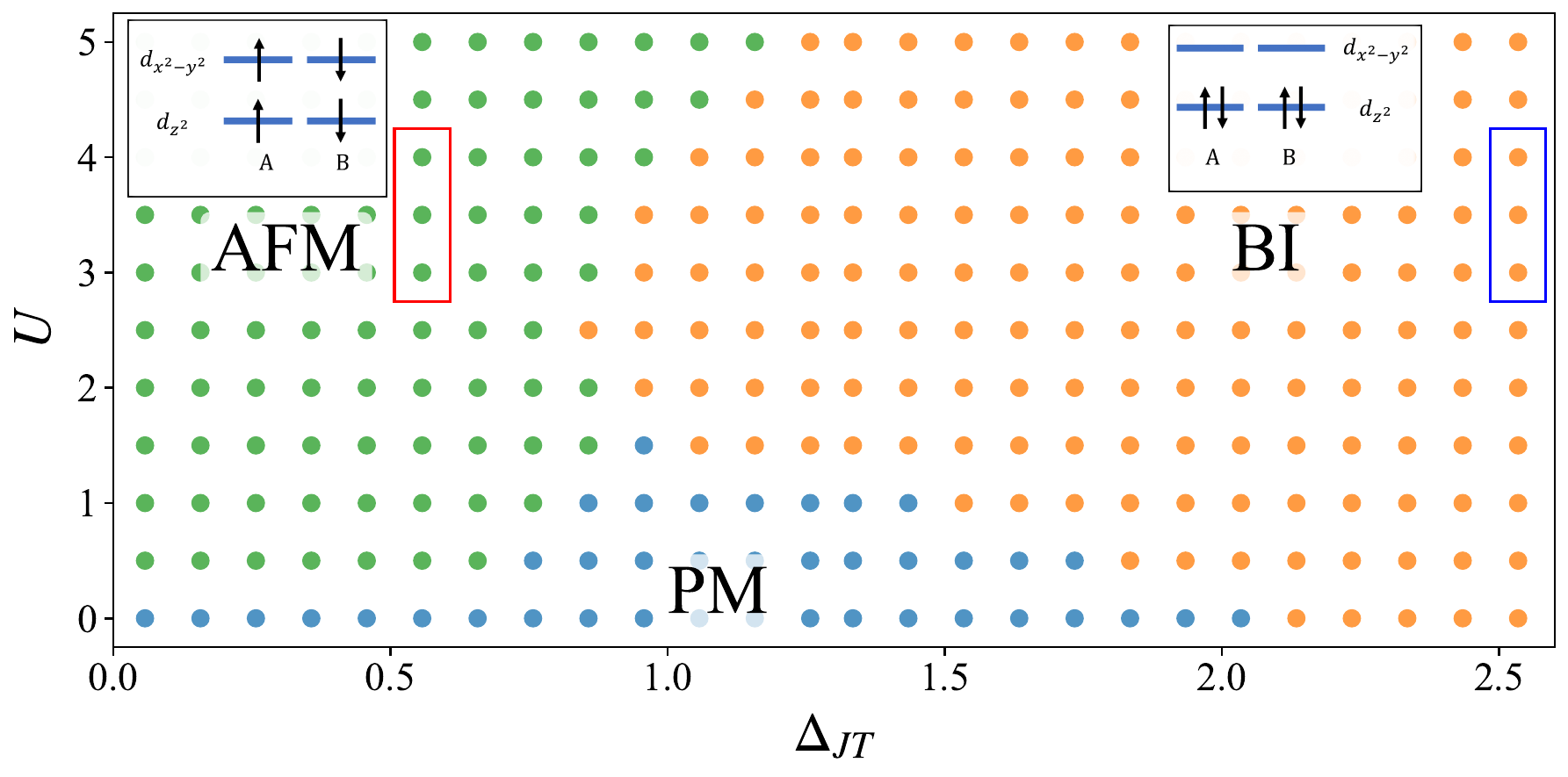} 
    \caption{Phase diagram of the isolated \ce{La2NiO4} layer obtained from the RISB calculation. The horizontal axis denotes the orbital splitting $\Delta_{JT}$, and the vertical axis is the Hubbard interaction strength $U$. The blue, green, and orange regions correspond to PM, AFM, and BI, respectively. 
    The upper inset schematically illustrates the AFM (right) and BI (left) electronic configurations.
    The red and blue boxes mark the values of $\Delta_{JT}$ extracted from the PBE and HSE06 parametrizations, respectively; their vertical extent corresponds to the estimated physically relevant interaction range $U\sim 3\text{--}4$ eV for the Ni $3d$ orbitals. These boxes indicate where the embedded 214 layer in \ce{La5Ni3O11} lies within the correlated phase diagram.}
    \label{fig:214_phase_diagramm}
\end{figure}

%\section{Results}
We can look at the phase diagram of the isolated 214 layer obtained within the RISB approach, shown in Fig.~\ref{fig:214_phase_diagramm} as a function of the Hubbard interaction $U$ and the Jahn--Teller splitting $\Delta_{JT}$. Since the $\mathrm{Ni}^{2+}$ ion hosts two $e_g$ electrons, the system naturally exhibits two competing insulating limits in addition to the weak-coupling paramagnetic metal, as illustrated schematically in Fig.~\ref{fig:214_phase_diagramm}.

In the interaction-dominated regime $U\gg\Delta_{JT}$, the $d_{x^2-y^2}$ and $d_{z^2}$ orbitals are nearly degenerate and each becomes singly occupied. Hund's coupling aligns the two spins into a local $S=1$ moment, which subsequently orders antiferromagnetically, giving rise to a correlation-driven AFM insulating state. In contrast, when the crystal-field splitting dominates $\Delta_{JT}\gg U$, both electrons occupy the lower-energy $d_{z^2}$ orbital while the $d_{x^2-y^2}$ orbital remains empty. The system is then driven into an orbital-polarized band-insulating state.

These two limits are connected by the RISB phase diagram shown in Fig.~\ref{fig:214_phase_diagramm}. At small $U$ and $\Delta_{JT}$, the system remains a PM. Increasing $U$ stabilizes the AFM phase with a finite staggered moment, whereas increasing $\Delta_{JT}$ suppresses magnetism and favors the BI phase through orbital polarization. The phase diagram therefore reveals two distinct mechanisms by which the intercalated 214 layer can be removed from low-energy metallic behavior: correlation-driven gap opening in the AFM phase and crystal-field-driven gap opening in the BI phase.

We now place the realistic parameters of the intercalated 214 layer onto the RISB phase diagram, as highlighted in Fig.~\ref{fig:214_phase_diagramm}. For the PBE parametrization, the Jahn--Teller splitting is relatively small $\Delta_{JT}=0.5571$ eV, placing the system in the AFM region for the physically relevant interaction strength $U\sim 3\text{--}4$ eV~\cite{327_ARPES_01,327_ARPES_02,327_DMFT_01}. At $U=4$ eV, the calculated staggered moment reaches $M=1.88~\mu_B$, consistent with a high-spin $S=1$ antiferromagnetic insulating state. In contrast, the HSE06 parametrization yields a much larger crystal-field splitting, $\Delta_{JT}=2.5341$ eV, which places the system deep inside the BI region. In this regime, the large orbital splitting overcomes Hund's coupling, driving the system into an orbital-polarized, nonmagnetic insulating state.

Although the PBE and HSE06 functionals favor different insulating phases, they lead to the same physical conclusion: the intercalated 214 layer is removed from the low-energy metallic sector. The PBE result represents the correlation-driven limit, whereas the HSE06 result represents the crystal-field-driven limit, demonstrating that the electronic inactivity of the 214 layer is robust against the uncertainty in the underlying density-functional description.

Finally, we examine whether coupling to the neighboring 327 blocks can electronically reactivate the intercalated 214 layer. 
Intuitively, if the 214 layer were metallic, inter-block hybridization would facilitate charge transfer between the 327 and 214 blocks, allowing 214-derived states to participate in the low-energy electronic structure. In contrast, if the 214 layer remains gapped, charge transfer is strongly suppressed, and its contribution to the Fermi-level states should remain negligible. To distinguish between these two possibilities, we construct a minimal coupled-layer model for \ce{La5Ni3O11}. The Hamiltonian of the 327 block is described in Appendix~\ref{Sec:A3} , following the established bilayer two-orbital description of high-pressure \ce{La3Ni2O7}~\cite{327_DFT_01,327_DFT_02,327_FLEX_01}.
Then, the effective Hamiltonian for \ce{La5Ni3O11} is
\begin{equation}
\label{eq:5311_Hamiltonian_01}
H_{\mathrm{eff}}^{5311}
=
H^{327}
+
H^{214}
+
H_{\perp},
\end{equation}
where $H^{327}$ and  $H^{214}$  describe the 327 and 214 blocks, respectively. The dominant interblock coupling arises from hopping between the $d_{z^2}$ orbitals of the intercalated 214 layer and the adjacent 327 block. Accordingly, the hybridization term takes the form
\begin{equation}
    \label{eq:5311_Hamiltonian_02}
    H_{\perp}
    =
    4t^{zz}_\perp\sum_{{\bm k},\sigma}
    \cos\frac{k_x}{2}\cos\frac{k_y}{2}
    \left(
    d_{{\bm k},z,\sigma}^{\dagger} c_{{\bm k},z,\sigma}
    +\mathrm{h.c.}
    \right)
\end{equation}
where $d_{{\bm k},z,\sigma}^{\dagger}$ and $c_{{\bm k},z,\sigma}^{\dagger}$ create electrons in the $d_{z^2}$ orbitals of the 214 layer and the neighboring 327 block, respectively. The hopping amplitude is $t_{\perp}^{zz}=-0.0237$ eV. This minimal model allows us to directly assess whether realistic interblock hybridization can generate 214-derived electronic states at the Fermi level.

%spectral weight back to the Fermi level. Figure~\ref{fig:energy_band} provides a direct layer-resolved test of this possibility using the minimal 327+214 model.

\begin{figure}
    \centering
    \includegraphics[width=\columnwidth]{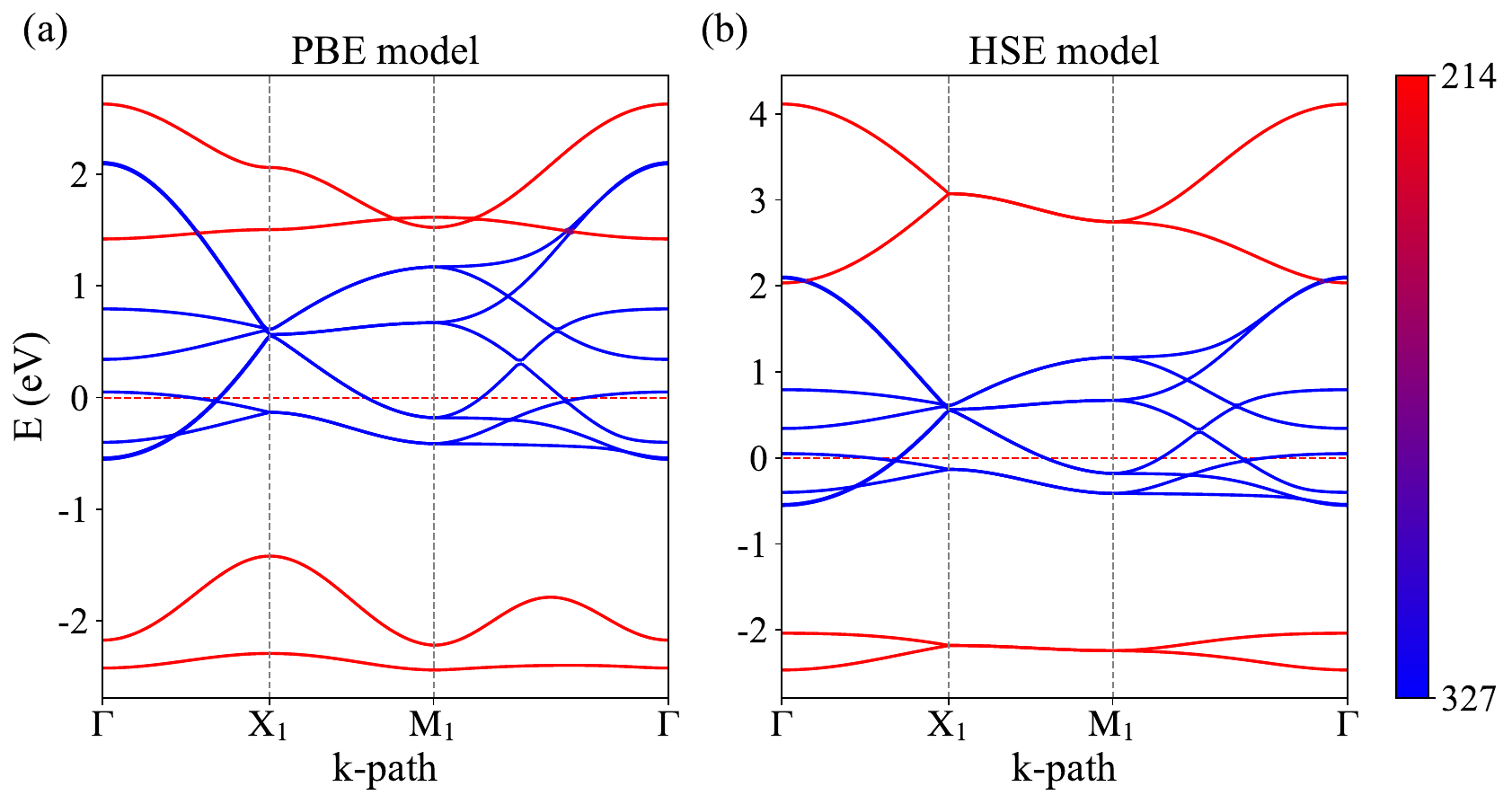} 
    \caption{Layer-resolved energy bands of the minimal 327+214 coupled-layer model for \ce{La5Ni3O11}: (a) PBE-derived 214 parameters and (b) HSE-derived 214 parameters. The color scale indicates the layer weight of each band, with blue denoting the 327 block and red denoting the 214 layer. In both cases, the bands near the Fermi level are predominantly 327-derived, and no additional 214-derived Fermi-level crossing appears.}
    \label{fig:energy_band}
\end{figure}

Using the same RISB framework, we solve the coupled Hamiltonian $H_{\mathrm{eff}}^{5311}$ self-consistently and obtain the renormalized band structure shown in Fig.~\ref{fig:energy_band}. The color scale indicates the layer-resolved spectral weight, with blue and red representing the 327 and 214 blocks, respectively. For both the PBE- and HSE-derived parametrizations, the bands crossing the Fermi level are overwhelmingly dominated by the 327 block, while the 214-derived bands remain well separated from $E_F$. No additional 214-derived Fermi-surface pocket emerges, demonstrating that the realistic $d_{z^2}$-$d_{z^2}$ interlayer hybridization is insufficient to reactivate the intercalated 214 layer. Although the high-energy electronic structure differs between the two parametrizations, the low-energy electronic structure is essentially identical.

Our results therefore establish that the intercalated \ce{La2NiO4} layer is electronically inactive at low energies and that the electronic structure of \ce{La5Ni3O11} is governed almost entirely by the \ce{La3Ni2O7} block. Among the two insulating scenarios identified for the isolated 214 layer, the band-insulating solution appears more compatible with the experimentally observed similarity between \ce{La5Ni3O11} and \ce{La3Ni2O7}, as it avoids introducing an additional magnetic subsystem between neighboring 327 blocks.

%Using the same strategy above, we calculate the $H_{\mathrm{eff}}^{5311}$ self-consistently and plot its band structure in Figure~\ref{fig:energy_band}.
%The color scale in Fig.~\ref{fig:energy_band} represents the relative layer weight of each band, with blue denoting 327 character and red denoting 214 character. In both the PBE- and HSE-derived parametrizations, the bands crossing or approaching $E_F$ are overwhelmingly blue, showing that the low-energy states remain dominated by the 327 block. The 214-derived spectral weight stays mainly away from $E_F$, and no additional 214-derived Fermi-level crossing is generated. Thus, the realistic $d_{z^2}$-$d_{z^2}$ interlayer hopping is insufficient to overcome the gapped character of the 214 layer. While the 214-related bands away from $E_F$ differ between the two parametrizations, the low-energy conclusion is unchanged.

%These results identify the 214 layer as an electronically inactive intercalated layer rather than an active low-energy component of \ce{La5Ni3O11}. Between the two gapped possibilities, the BI interpretation appears more natural for the full 5311 material: a magnetic 214 layer would insert an additional ordered component between neighboring 327 blocks, whereas a nonmagnetic gapped layer is more consistent with the observed similarity between 5311 and bulk 327.

\section{Discussion and conclusion}

The hybrid structure of \ce{La5Ni3O11} might suggest a low-energy electronic structure built from both 327 and 214 constituents. Our results instead support a more selective picture. In the isolated 214-layer model, realistic PBE- and HSE-derived parameters both place the system outside the paramagnetic metallic regime. Although the two functionals favor different microscopic limits---a Hund-stabilized high spin AFM state for the smaller PBE orbital splitting and an orbital-polarized BI state for the larger HES06 splitting---both limits remove the 214 layer from the low-energy metallic sector.

The coupled-layer calculation shows that this conclusion is not overturned by hybridization with the neighboring \ce{La3Ni2O7} block. The realistic $d_{z^2}$-$d_{z^2}$ interlayer hopping does not generate an additional 214-derived Fermi-level crossing, and the bands near $E_F$ remain dominated by the 327 block. Thus, the 214 layer does not act as an independent source of low-energy carriers in \ce{La5Ni3O11}.

This conclusion also clarifies the distinction between \ce{La5Ni3O11} and bulk \ce{La3Ni2O7}. The low-energy electronic degrees of freedom in both systems originate primarily from the 327 block, explaining their similar pressure-induced phenomenology. However, in \ce{La5Ni3O11} the electronically inactive 214 layer is intercalated between neighboring 327 blocks. In this case, we find instead of contributing an additional Fermi surface, this layer weakens coherent interblock hopping along the $c$ axis and makes the 327-derived electronic structure more quasi-two-dimensional than in bulk 327.
This role of the intercalated layer is reminiscent of FeSe-based intercalated superconductors, where electronically active FeSe layers are separated by alkali-metal, hydroxide, or molecular spacer layers that tune charge transfer, layer spacing, and dimensionality without necessarily providing the primary Fermi-surface states~\cite{FeSe_intercalation_01,FeSe_intercalation_02}. The analogy should not be interpreted as implying the same pairing mechanism. Rather, it highlights a broader structural principle: intercalation can reshape the dimensionality and interlayer coherence of an active superconducting block without adding an independent low-energy electronic subsystem.

Consequently, the minimal low-energy description of \ce{La5Ni3O11} is governed primarily by the \ce{La3Ni2O7} block, with the intercalated \ce{La2NiO4} layer remaining electronically inactive. From this perspective, \ce{La5Ni3O11} can be viewed as a structurally modulated, more two-dimensional realization of the 327 electronic system, rather than a distinct multiblock electronic material.

Our prediction can be tested directly by angle-resolved photoemission spectroscopy (ARPES). If the intercalated 214 layer is indeed electronically inactive, the Fermi surface of \ce{La5Ni3O11} should closely resemble that of \ce{La3Ni2O7}, without additional 214-derived Fermi pockets. Encouragingly, recent ARPES measurements on \ce{La5Ni3O11} thin films are already consistent with this picture, revealing low-energy Fermi-surface contours that closely match those of \ce{La3Ni2O7} ~\cite{327_ARPES_03}. Our work therefore provides a unified low-energy description of \ce{La5Ni3O11} and clarifies the electronic role of the intercalated \ce{La2NiO4} layer in superconducting Ruddlesden--Popper nickelates.

%Consequently, the minimal low-energy description of \ce{La5Ni3O11} should be based primarily on the \ce{La3Ni2O7} block, with the \ce{La2NiO4} layer treated as an electronically inactive intercalated layer. In this sense, \ce{La5Ni3O11} may be viewed as a structurally modulated, more two-dimensional realization of the 327 electronic system.
%Experimentally, our conclusion can be verified directly from angle-resolved photoemission spectroscopy (ARPES), by comparing \ce{La5Ni3O11} and \ce{La3Ni2O7} Fermi surfaces. Recent ARPES in \ce{La5Ni3O11} thin film already supports our conclusion, where the low-energy FSs host the same contour as \ce{La3Ni2O7}.
%We hope our results provide a new perspective on this intercalated structure in nickelates.

\section{Acknowledgement}
We acknowledge the support by the National Natural Science Foundation of China (Grant NSFC-12494594, NSFC-12574150), the Chinese Academy of Sciences Project for Young Scientists in Basic Research (2022YSBR-048), the New Cornerstone Investigator Program.

\appendix
\section{Rotationally invariant slave-boson formalism}
\label{Sec:A1}

In this appendix, we briefly summarize the rotationally invariant slave-boson (RISB) formalism used to solve the two-orbital model introduced in the main text. For notational simplicity, the two orbitals are labeled by \(a=1,2\), corresponding to the \(d_{x^2-y^2}\) and \(d_{z^2}\) orbitals, respectively. The interaction parameters follow the same conventions as in the main text, with \(U'=U-2J_H\) and \(J_C=J_H\).

For the two-orbital Hamiltonian with the local interaction term given in Eq.~\ref{eq:214_Hamiltonian_02}, the local Hilbert space consists of 16 atomic eigenstates \(|\Gamma\rangle\). In the rotationally invariant formulation, the slave-boson amplitudes \(\phi_{\Gamma n}\) connect the local interacting eigenstates \(|\Gamma\rangle\) with quasiparticle Fock states \(|n\rangle\). As a result, the number of independent slave-boson amplitudes is larger than the number of local eigenstates; for the present two-orbital case, this leads to a total of 20 slave bosons. The local eigenstates, their quantum numbers, and the corresponding slave bosons are listed in Table~\ref{tab:eigenstates_01}.

\begin{table*}[htbp]
    \begin{center}
        \caption{Eigenstates \(|\Gamma\rangle\) of the \(SU(2)\) rotationally invariant two-orbital Hubbard model~\cite{RISB_01,RISB_02}. The spin quantum numbers and local energies are given for each state. The last column lists the corresponding slave-boson amplitudes used in the RISB formalism.}
        \label{tab:eigenstates_01}
        \setlength{\tabcolsep}{8pt}
        \begin{tabular}{c c c c c c }
            \hline
            \hline
            No. 
            & Eigenstate \(|\Gamma\rangle\) & \(S_{\Gamma}\) & \(S_{\Gamma}^{z}\) & \(E_{\Gamma}\) & \(\phi_{\Gamma n}\) \\
            \hline
            1 & \(|00,00\rangle\) & 0 & 0 & 0 & \(\phi_{1,1}\) \\
            \hline
            2 & \(|\uparrow0,00\rangle\) & \(\frac{1}{2}\) & \(\frac{1}{2}\) & 0 & \(\phi_{2,2}\) \\
            3 & \(|0\downarrow,00\rangle\) & \(\frac{1}{2}\) & \(-\frac{1}{2}\) & 0 & \(\phi_{3,3}\) \\
            4 & \(|00,\uparrow0\rangle\) & \(\frac{1}{2}\) & \(\frac{1}{2}\) & 0 & \(\phi_{4,4}\) \\
            5 & \(|00,0\downarrow\rangle\) & \(\frac{1}{2}\) & \(-\frac{1}{2}\) & 0 & \(\phi_{5,5}\) \\
            \hline
            6 & \(|\uparrow0,\uparrow0\rangle\) & 1 & 1 & \(U'-J_H\) & \(\phi_{6,8}\) \\
            7 & \(\frac{1}{\sqrt{2}}\left( |\uparrow0,0\downarrow\rangle + |0\downarrow,\uparrow0\rangle \right)\) & 1 & 0 & \(U'-J_H\) & \( (\phi_{7,10}, \phi_{7,11})\) \\
            8 & \(|0\downarrow,0\downarrow\rangle\) & 1 & \(-1\) & \(U'-J_H\) & \(\phi_{8,9}\) \\
            9 & \(\frac{1}{\sqrt{2}}\left(|\uparrow0,0\downarrow\rangle - |0\downarrow,\uparrow0\rangle\right)\) & 0 & 0 & \(U'+J_H\) &  \( (\phi_{9,10}, \phi_{9,11}) \) \\
            10 & \(\frac{1}{\sqrt{2}}\left( |\uparrow\downarrow,00\rangle - |00,\uparrow\downarrow\rangle \right)\) & 0 & 0 & \(U-J_C\) & \( (\phi_{10,6}, \phi_{10,7})\)\\
            11 & \(\frac{1}{\sqrt{2}}\left(|\uparrow\downarrow,00\rangle + |00,\uparrow\downarrow\rangle\right)\) & 0 & 0 & \(U+J_C\) & \( (\phi_{11,6}, \phi_{11,7})\) \\
            \hline
            12 & \(|\uparrow\downarrow,\uparrow0\rangle\) & \(\frac{1}{2}\) & \(\frac{1}{2}\) & \(U+2U'-J_H\) & \(\phi_{12,12}\) \\
            13 & \(|\uparrow\downarrow,0\downarrow\rangle\) & \(\frac{1}{2}\) & \(-\frac{1}{2}\) & \(U+2U'-J_H\) & \(\phi_{13,13}\) \\
            14 & \(|\uparrow0,\uparrow\downarrow\rangle\) & \(\frac{1}{2}\) & \(\frac{1}{2}\) & \(U+2U'-J_H\) & \(\phi_{14,14}\) \\
            15 & \(|0\downarrow,\uparrow\downarrow\rangle\) & \(\frac{1}{2}\) & \(-\frac{1}{2}\) & \(U+2U'-J_H\) & \(\phi_{15,15}\) \\
            \hline
            16 & \(|\uparrow\downarrow,\uparrow\downarrow\rangle\) & 0 & 0 & \(2U+4U'-2J_H\) & \(\phi_{16,16}\) \\
            \hline
            \hline
        \end{tabular}
    \end{center}
\end{table*}

The general two-orbital Hamiltonian can be written as
\begin{equation}
    \label{eq:appendix_H_01}
    \begin{aligned}
        H &= H_{\mathrm{kin}} + H_{\mathrm{loc}}, \\
        H_{\mathrm{kin}} &= \sum_{a=1,2} \sum_{{\bm k},\sigma} \epsilon_a\, d_{{\bm k}a\sigma}^\dagger d_{{\bm k}a\sigma} \\
        &\quad+ \sum_{a,b=1,2} \sum_{{\bm k},\sigma} \varepsilon_{ab}({\bm k})\, d_{{\bm k}a\sigma}^\dagger d_{{\bm k}b\sigma}, \\
        H_{\mathrm{loc}} &= U \sum_i \sum_{a=1,2} n_{ia\uparrow}n_{ia\downarrow} \\
        &\quad + U' \sum_i \sum_{\sigma\sigma'} n_{i1\sigma}n_{i2\sigma'} \\
        &\quad - J_H \sum_i \sum_{\sigma} n_{i1\sigma}n_{i2\sigma} \\
        &\quad + J_H \sum_i \sum_{\sigma} d_{i1\sigma}^\dagger d_{i2\bar{\sigma}}^\dagger d_{i1\bar{\sigma}} d_{i2\sigma} \\
        &\quad + J_C \sum_i \left(
        d_{i1\uparrow}^\dagger d_{i1\downarrow}^\dagger d_{i2\downarrow} d_{i2\uparrow}
        + \text{H.c.}
        \right),
    \end{aligned}
\end{equation}
where \(d_{{\bm k}a\sigma}^\dagger\) (\(d_{{\bm k}a\sigma}\)) creates (annihilates) an electron with momentum \({\bm k}\), orbital \(a\), and spin \(\sigma\), and \(n_{ia\sigma}=d_{ia\sigma}^\dagger d_{ia\sigma}\). The noninteracting part \(H_{\mathrm{kin}}\) contains the orbital onsite energies and kinetic terms, whereas \(H_{\mathrm{loc}}\) describes the local Coulomb interaction, Hund's exchange, and pair-hopping processes.

In the RISB formalism, the physical electron operators are represented in terms of quasiparticle fermions \(f_{a\sigma}\) and slave bosons through a renormalization matrix \(R\),
\begin{equation}
    d_{a\sigma} = \sum_b R_{ab,\sigma}\, f_{b\sigma},
\end{equation}
where \(R\) is determined self-consistently from the condensed slave-boson amplitudes. In the slave-boson basis adopted in the present work, \(R\) is taken to be diagonal in orbital space, so that only the orbital-diagonal quasiparticle renormalization factors are retained. With this mapping, the Hamiltonian in the enlarged Hilbert space takes the form
\begin{equation}
    \label{eq:appendix_H_02}
    \begin{aligned}
        \underline{H} &= \underline{H}_{\mathrm{kin}} + \underline{H}_{\mathrm{loc}}, \\
        \underline{H}_{\mathrm{kin}} &= \sum_{{\bm k},\sigma} \sum_{a,b=1,2}
        \left[ R_\sigma^\dagger\, \varepsilon({\bm k})\, R_\sigma \right]_{ab}
        f_{{\bm k}a\sigma}^\dagger f_{{\bm k}b\sigma} \\
        &\quad+ \sum_{{\bm k},\sigma}\sum_{a=1,2} \epsilon_a\, f_{{\bm k}a\sigma}^\dagger f_{{\bm k}a\sigma}, \\
        \underline{H}_{\mathrm{loc}} &= \sum_{i,\Gamma,n} E_\Gamma\, \phi_{i\Gamma n}^\dagger \phi_{i\Gamma n}.
    \end{aligned}
\end{equation}
In this representation, the interaction part is diagonal in the local eigenbasis, whereas the kinetic term is renormalized by the \(R\) matrices.

Since the slave-boson construction enlarges the Hilbert space, constraints must be imposed to recover the physical subspace. The first constraint is the normalization of the slave-boson amplitudes,
\begin{equation}
    \label{eq:appendix_constraint_condition_01}
    \sum_{\Gamma n} \phi_{\Gamma n}^\dagger \phi_{\Gamma n} = 1.
\end{equation}
The second constraint relates the quasiparticle density to the slave-boson amplitudes. In the present implementation, we work in a basis where the off-diagonal quasiparticle density-matrix elements vanish, so that only the diagonal constraints are retained:
\begin{equation}
    \label{eq:appendix_constraint_condition_02}
    \sum_{\Gamma n} \phi_{\Gamma n}^\dagger \phi_{\Gamma n}
    \langle n| f_{a\sigma}^\dagger f_{a\sigma} |n\rangle
    =
    f_{a\sigma}^\dagger f_{a\sigma}.
\end{equation}

These constraints are enforced by introducing Lagrange multipliers, with \(\lambda_0\) corresponding to the normalization condition and \(\Lambda_{a\sigma}\) to the density constraints. Within the saddle-point approximation, the slave-boson operators are replaced by their condensed values, \(\langle \phi_{\Gamma n}\rangle=\varphi_{\Gamma n}\), and the free energy can be written as
\begin{equation}
    \label{eq:appendix_free_energy_01}
    \begin{aligned}
        F =& -\frac{1}{\beta}\sum_{{\bm k},j}\ln\!\left(1+e^{-\beta \varepsilon_{{\bm k},j}}\right) - \lambda_0 N \\
        &+ N \sum_{\Gamma n} \varphi_{\Gamma n}^2
        \left[
        E_\Gamma + \lambda_0
        -\sum_{a\sigma}\Lambda_{a\sigma}
        \langle n|f_{a\sigma}^\dagger f_{a\sigma}|n\rangle
        \right],
    \end{aligned}
\end{equation}
where \(\varepsilon_{{\bm k},j}\) are the quasiparticle band energies and \(N\) is the total number of lattice sites. The saddle-point solution is obtained by minimizing the free-energy functional with respect to the variables \(\{\varphi_{\Gamma n},\lambda_0,\Lambda_{a\sigma}\}\).

At \(T=0\), the corresponding ground-state energy is
\begin{equation}
    \label{eq:appendix_ground_energy_01}
    \begin{aligned}
        E =& \sum_{{\bm k},j} \varepsilon_{{\bm k},j}\, n_F(\varepsilon_{{\bm k},j}) \\
        &+ N \sum_{\Gamma n} \varphi_{\Gamma n}^2
        \left[
        E_\Gamma
        -\sum_{a\sigma}\Lambda_{a\sigma}
        \langle n|f_{a\sigma}^\dagger f_{a\sigma}|n\rangle
        \right].
    \end{aligned}
\end{equation}

In the antiferromagnetic calculations discussed in the main text, the above saddle-point equations are solved within a two-sublattice (\(A,B\)) scheme. Once the saddle-point solution is obtained, the quasiparticle renormalization, orbital occupations, local magnetic moments, and the low-energy band structure can be evaluated straightforwardly.

\section{Refined analysis of the AFM sector}
\label{Sec:A2}
In the main text, the AFM phase is classified solely according to the presence of magnetic order. Here we further refine the AFM sector and distinguish between low-spin and high-spin antiferromagnetic solutions in the parameter region relevant to the 214 layer.

\subsection{Phase diagram near the phase-transition boundary}
\label{Sec:A2.1}

\begin{figure}[htbp]
    \centering
    \includegraphics[width=1.0\linewidth]{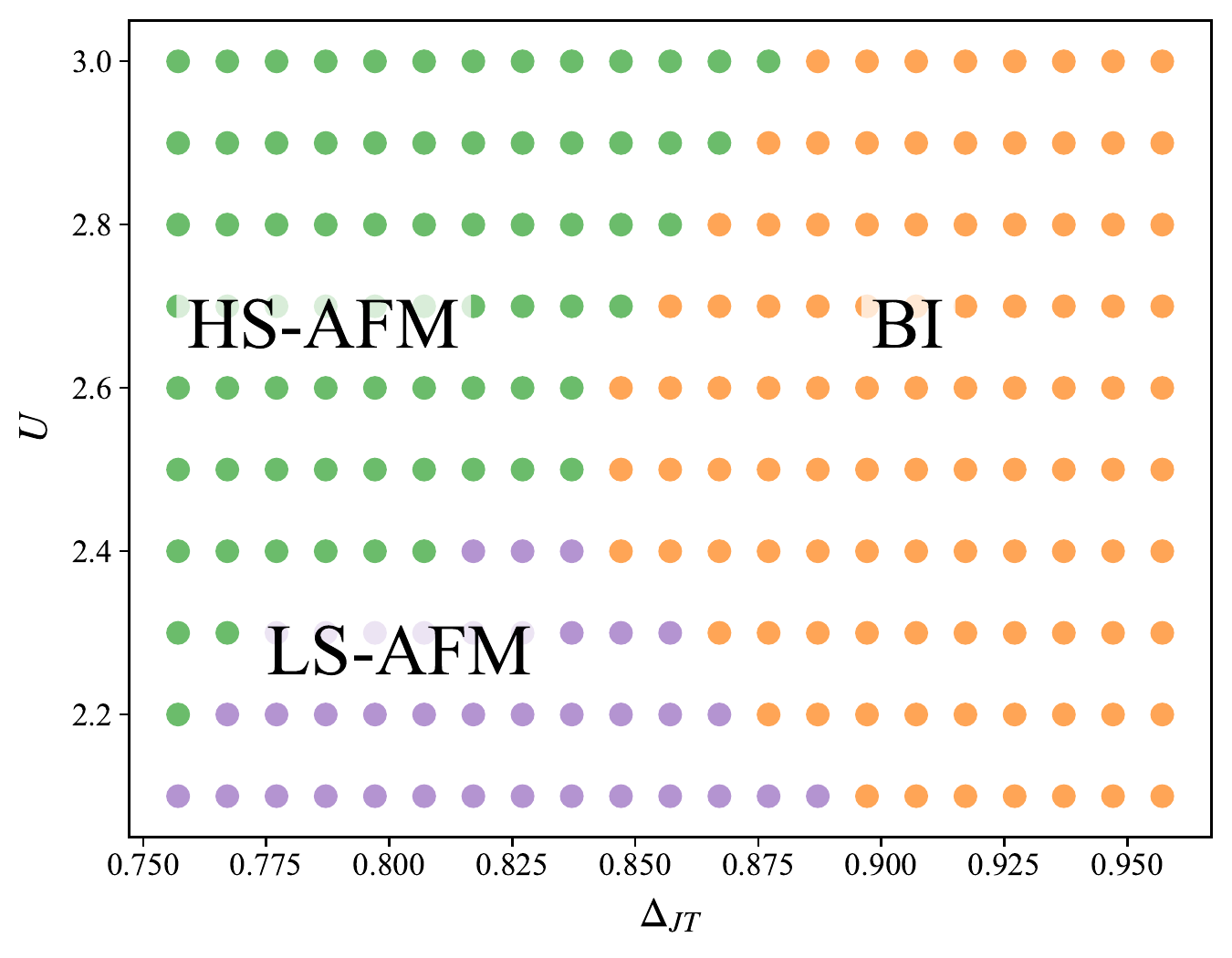} 
    \caption{Refined electronic phase diagram of the isolated \ce{La2NiO4} layer obtained from the RISB calculation in the parameter window near the phase boundaries of Fig.~\ref{fig:214_phase_diagramm}. The horizontal axis denotes the orbital splitting $\Delta_{\mathrm{JT}}$, and the vertical axis is the local Coulomb interaction strength $U$. Unless otherwise noted, the remaining parameters are the same as those used in Fig.~\ref{fig:214_phase_diagramm}. The green, purple, and orange regions correspond to HS-AFM, LS-AFM, and BI, respectively.}
    \label{fig:appendix_214_phase_diagramm}
\end{figure}

In Fig.~\ref{fig:appendix_214_phase_diagramm}, we focus on the parameter window $U=2\text{--}3$ eV and $\Delta_{\mathrm{JT}}=0.755\text{--}0.955$ eV in order to resolve the fine structure of the magnetic sector. A key feature of this phase diagram is that the AFM region can be further divided into two distinct parts, namely the high-spin antiferromagnetic (HS-AFM) state and the low-spin antiferromagnetic (LS-AFM) state. The distinction between them is reflected mainly in the orbital occupancies and in the magnitude of the local magnetic moment.

The HS-AFM state appears at relatively larger $U$, where the occupancies of the two orbitals become close to a $1{:}1$ ratio and the local moment is strongly enhanced. By contrast, the LS-AFM state occurs when $U$ is not yet large enough to fully compensate the effect of orbital splitting. In this regime, magnetic order is already present, but the orbital occupancies remain substantially imbalanced and the total local magnetic moment stays below $1\mu_B$. In this sense, the LS-AFM state may be viewed as an intermediate magnetic regime between the PM state and the HS-AFM state.

Another notable feature is that the BI phase appears near the boundary between the LS-AFM and HS-AFM regions. This indicates that the competition between the interaction strength $U$ and the orbital splitting $\Delta_{\mathrm{JT}}$ makes the orbital occupancy distribution unstable in this parameter range. As a result, a nonmagnetic BI solution can become stabilized between the two magnetic states.

The distinction between LS-AFM and HS-AFM itself is closely related to the broader competition between low-spin and high-spin states that has been discussed in previous theoretical studies of other nickelate systems~\cite{5410_magnetic_structure_01}. As will be shown more explicitly in Sec.~\ref{Sec:A2.3}, the LS-AFM and HS-AFM states also differ qualitatively in their low-energy electronic structures.

\subsection{Order parameters along representative \texorpdfstring{$\Delta_{JT}$}{ΔJT} cuts}
\label{Sec:A2.2}

This subsection examines the evolution of representative order parameters across the phase transitions in order to further characterize the phases identified in Fig.~\ref{fig:appendix_214_phase_diagramm}. We focus on the orbital-resolved electron occupation $N$ and the local magnetic moment $M$. For AFM solutions, the local moments on the two sublattices have equal magnitude and opposite sign, $M_A=-M_B$, whereas for PM and BI both vanish, $M_A=M_B=0$. In the following, we therefore show only the results for sublattice $A$.

\begin{figure}[htbp]
    \centering
    \includegraphics[width=1.0\linewidth]{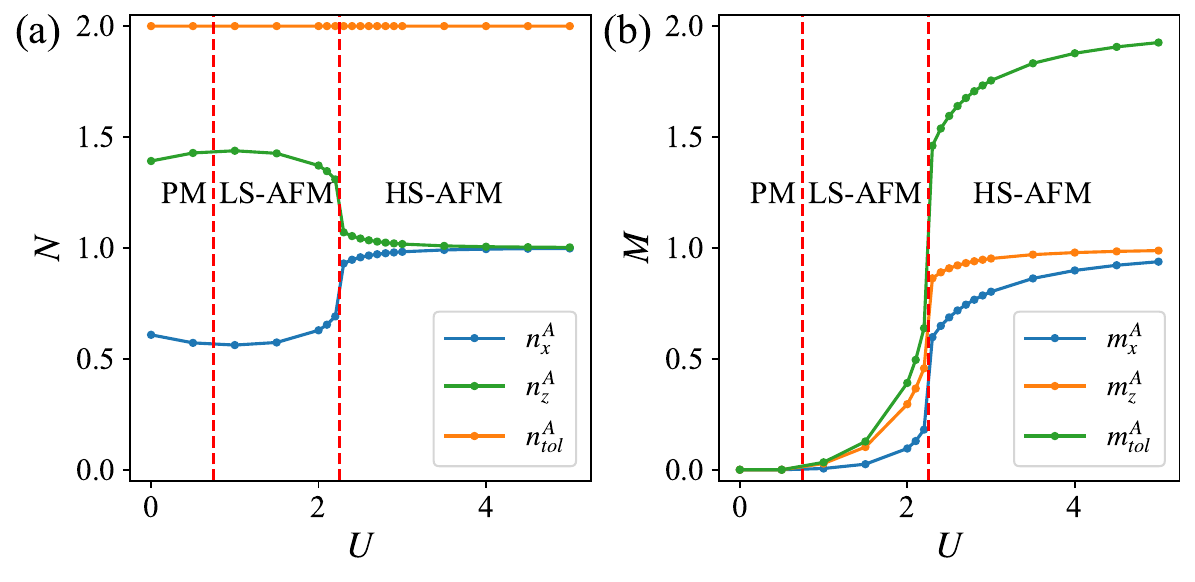} 
    \caption{Evolution of the orbital-resolved electron occupation $N$ and local magnetic moment $M$ as functions of $U$, obtained from the RISB calculation at fixed $\Delta_{\mathrm{JT}} = 0.7571$ eV. Panels (a) and (b) show the $U$-dependence of $N$ and $M$ on sublattice $A$, respectively. The red dashed lines indicate the phase boundaries.}
    \label{fig:appendix_214_orderparameter_01}
\end{figure}

Figure~\ref{fig:appendix_214_orderparameter_01} shows the order parameters at fixed $\Delta_{\mathrm{JT}} = 0.7571$ eV. Along this cut, the system evolves from PM to LS-AFM and then to HS-AFM as $U$  increases. As shown in Fig.~\ref{fig:appendix_214_orderparameter_01}(a), the PM-to-LS-AFM transition does not involve a pronounced redistribution of the orbital occupancies: the $d_{x^2-y^2}$ and $d_{z^2}$ occupations remain relatively close to those in the PM state, although a finite magnetic order has already developed. In the LS-AFM regime, the orbital occupancies are still strongly imbalanced, with the $d_{x^2-y^2}$ orbital occupied by about $0.6$ electrons and the $d_{z^2}$ orbital by about $1.4$ electrons.

The corresponding evolution of the local magnetic moments is shown in Fig.~\ref{fig:appendix_214_orderparameter_01}(b). In the PM phase, the local moments vanish. Upon entering the LS-AFM phase, finite but relatively small moments develop, and the total local magnetic moment remains below $1\mu_B$. As $U$ increases further, the system undergoes a transition from LS-AFM to HS-AFM. This transition is accompanied by a clear redistribution of the orbital occupancies toward a nearly $1{:}1$ configuration and by a simultaneous jump in the magnetic moment, indicating a first-order character in the charge and magnetic sectors. In the HS-AFM phase, the moments on both orbitals become much larger than in the LS-AFM regime.

\begin{figure}[htbp]
    \centering
    \includegraphics[width=1.0\linewidth]{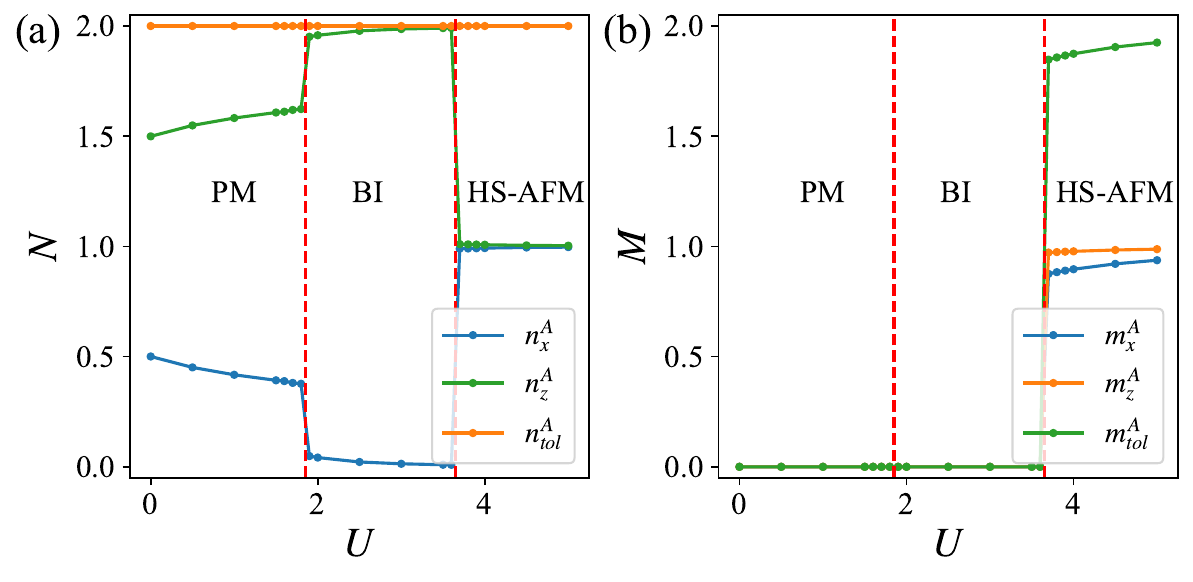} 
    \caption{Evolution of the orbital-resolved electron occupation $N$ and local magnetic moment $M$ as functions of $U$, obtained from the RISB calculation at fixed $\Delta_{\mathrm{JT}} = 0.9571$ eV. All other parameters are the same as in Fig.~\ref{fig:appendix_214_orderparameter_01}.}
    \label{fig:appendix_214_orderparameter_02}
\end{figure}

Figure~\ref{fig:appendix_214_orderparameter_02} shows the corresponding results at fixed $\Delta_{\mathrm{JT}} = 0.9571$ eV. In this case, the system evolves from PM to BI and then to HS-AFM as $U$ increases. As shown in Fig.~\ref{fig:appendix_214_orderparameter_02}(a), the BI phase is characterized by very strong orbital polarization: the occupation of the $d_{z^2}$ orbital approaches $2$, while that of the $d_{x^2-y^2}$ orbital approaches $0$. At the PM-to-BI phase boundary, the occupancies show a noticeable jump, signaling a sudden reconstruction of the electronic configuration. Figure~\ref{fig:appendix_214_orderparameter_02}(b) further shows that both PM and BI remain nonmagnetic, with vanishing local moments, whereas the HS-AFM phase is marked by a rapid increase of the local magnetic moment to values close to $2\mu_B$.

Taken together, the two representative $\Delta_{\mathrm{JT}}$ cuts provide a quantitative illustration of the phase classifications in Fig.~\ref{fig:appendix_214_phase_diagramm}. At smaller $\Delta_{\mathrm{JT}}$, increasing $U$ first stabilizes the LS-AFM phase before driving the system into the HS-AFM phase, whereas at larger $\Delta_{\mathrm{JT}}$ the system instead passes through a BI phase prior to entering the HS-AFM regime. These trends reflect the competition between orbital splitting and electron interaction discussed in Sec.~\ref{Sec:A2.1}.

\subsection{Representative band structures}
\label{Sec:A2.3}
\begin{figure*}[htbp]
    \centering
    \includegraphics[width=0.8\linewidth]{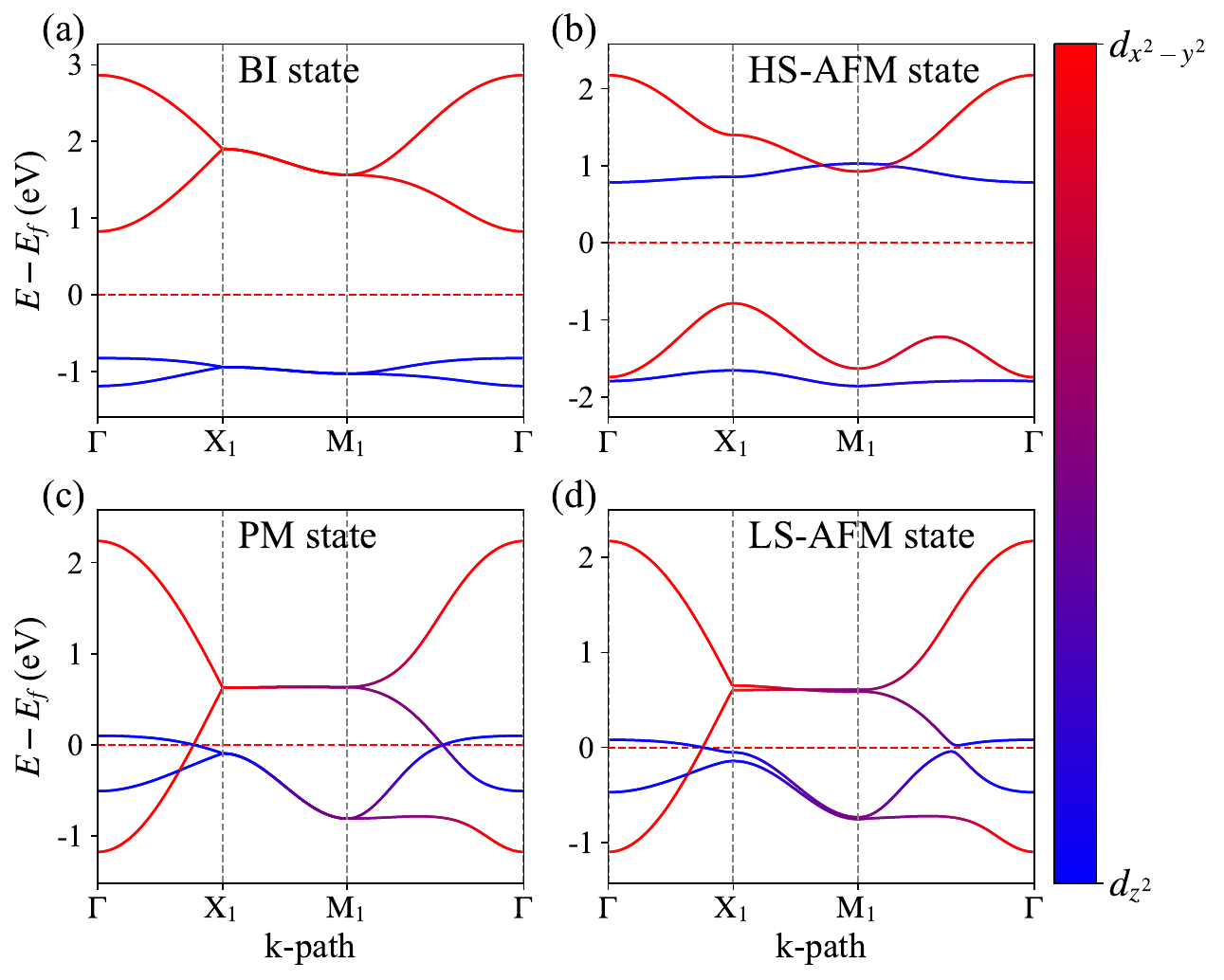} 
    \caption{Representative band structures of the isolated \ce{La2NiO4} layer in four typical phases: (a) BI, (b) HS-AFM, (c) PM, and (d) LS-AFM. The bands are plotted along the high-symmetry path of the folded Brillouin zone. The red-blue color scale indicates the orbital character of each band, with red corresponding to the $d_{x^2-y^2}$ orbital and blue to the $d_{z^2}$ orbital. The BI and HS-AFM states are gapped, the PM state is metallic, and the LS-AFM state is semimetallic.}
    \label{fig:energy_band_02}
\end{figure*}

To further clarify the low-energy electronic character of the phases identified in Fig.~\ref{fig:appendix_214_phase_diagramm}, we show in Fig.~\ref{fig:energy_band_02} the representative band structures of four typical states: BI, HS-AFM, PM, and LS-AFM. The corresponding parameter sets are chosen as follows: $U=3.0$, $J_H/U=0.1$, and $\Delta_{\mathrm{JT}}=0.9571$ eV for the BI state; $U=3.0$, $J_H/U=0.1$, and $\Delta_{\mathrm{JT}}=0.5571$ eV for the HS-AFM state; $U=0.0$ and $\Delta_{\mathrm{JT}}=0.5571$ eV for the PM state; and $U=1.0$, $J_H/U=0.1$, and $\Delta_{\mathrm{JT}}=0.5571$ eV for the LS-AFM state. The red-blue color scale indicates the relative orbital weight, with red and blue corresponding to the $d_{x^2-y^2}$ and $d_{z^2}$ orbitals, respectively.

The PM state, shown in Fig.~\ref{fig:energy_band_02}(c), exhibits clear band crossings at the Fermi level and is therefore metallic. By contrast, the LS-AFM state in Fig.~\ref{fig:energy_band_02}(d) is not fully gapped. Although antiferromagnetic order is already present, the low-energy bands still touch near the Fermi level, giving rise to a semimetallic state. This distinguishes the LS-AFM phase from a conventional insulating AFM state and places it between the PM and HS-AFM phases in terms of its low-energy electronic character.

The HS-AFM and BI states are both gapped, but their physical origins are different. As shown in Fig.~\ref{fig:energy_band_02}(b), the HS-AFM phase develops a full gap once the interaction strength is large enough to drive the system into the high-moment antiferromagnetic regime. The BI state in Fig.~\ref{fig:energy_band_02}(a) is also insulating, but its gap originates primarily from strong orbital polarization induced by the orbital splitting rather than by magnetic ordering.

These representative band structures are fully consistent with the phase classifications and order-parameter evolutions discussed in Secs.~\ref{Sec:A2.1} and \ref{Sec:A2.2}. In particular, they confirm that the LS-AFM phase is characterized by a small local magnetic moment together with a semimetallic low-energy band structure, whereas the HS-AFM phase combines a much larger local magnetic moment with a fully gapped spectrum. This difference justifies resolving the AFM sector into LS-AFM and HS-AFM in the appendix, while treating it as a single broader category in the main text.

\section{Low-energy model for the \texorpdfstring{\ce{La3Ni2O7}} block}
\label{Sec:A3}

In this appendix, we summarize the low-energy model adopted for the \ce{La3Ni2O7} block used in the minimal 327+214 coupled-layer calculation in the main text. Since the present work focuses on the electronic role of the intermediate \ce{La2NiO4} layer in \ce{La5Ni3O11}, we do not attempt to re-examine the full correlated phase diagram of \ce{La3Ni2O7}. Instead, we employ a DFT-based low-energy model for the \ce{La3Ni2O7} block and include the effect of local electron interactions at the RISB saddle-point level, while restricting the solution to the paramagnetic sector.

\begin{table}[htbp]
    \centering
    \caption{Tight-binding parameters of the \ce{La3Ni2O7} block used in the present work. All values are in eV.}
    \label{tab:327_hopping_parameters}
    \begin{tabular}{c c c c c c}
        \hline
        \hline
        $t_{1x}^{327}$ & $t_{1z}^{327}$ & $t_{2x}^{327}$ & $t_{2z}^{327}$ & $t_{\perp x}^{327}$ & $t_{\perp z}^{327}$ \\
        \hline
        -0.4334 & -0.0880 & 0.0712 & -0.0167 & 0.0078 & -0.5802 \\
        \hline
        & $t_{3xz}^{327}$ & $t_{4xz}^{327}$ & $\epsilon_x^{327}$ & $\epsilon_z^{327}$ & \\
        \hline
        & 0.2060 & -0.0287 & 0.5327 & 0.1810 & \\
        \hline
        \hline
    \end{tabular}
\end{table}

The low-energy Hamiltonian of the \ce{La3Ni2O7} block is written as
\begin{equation}
    \label{eq:appendix_327_H_01}
    H^{327}=H_{0}^{327}+H_{\mathrm{int}}^{327},
\end{equation}
where $H_{0}^{327}$ denotes the noninteracting tight-binding Hamiltonian extracted from the DFT fitting, and $H_{\mathrm{int}}^{327}$ is the local interaction term. The kinetic part contains the intra-layer, inter-layer, and inter-orbital hopping processes relevant to the bilayer \ce{La3Ni2O7} structure,
\begin{equation}
    \label{eq:appendix_327_H_02}
    \begin{aligned}
        H_{0}^{327} = & \sum_{{\bm k}, l, \sigma} \Big[ \epsilon_x^{327} + 2t_{1x}^{327} (\cos k_x + \cos k_y) \\
        &+ 4 t_{2x}^{327} \cos k_x \cos k_y \Big] c_{{\bm k},x,l,\sigma}^\dagger c_{{\bm k},x,l,\sigma} \\
        +& \sum_{{\bm k}, l, \sigma} \Big[ \epsilon_z^{327} + 2t_{1z}^{327} (\cos k_x + \cos k_y) \\
        &+ 4 t_{2z}^{327} \cos k_x \cos k_y \Big] c_{{\bm k},z,l,\sigma}^\dagger c_{{\bm k},z,l,\sigma} \\
        +& \sum_{{\bm k}, l, \sigma} 2t_{3xz}^{327} (\cos k_x - \cos k_y)
        \left(c_{{\bm k},x,l,\sigma}^\dagger c_{{\bm k},z,l,\sigma} + \mathrm{H.c.}\right) \\
        +& \sum_{{\bm k}, \sigma} 2t_{4xz}^{327} (\cos k_x - \cos k_y) \\
        &\times \left(c_{{\bm k},x,t,\sigma}^\dagger c_{{\bm k},z,b,\sigma} + c_{{\bm k},x,b,\sigma}^\dagger c_{{\bm k},z,t,\sigma} + \mathrm{H.c.}\right) \\
        +& \sum_{{\bm k}, \sigma} t_{\perp x}^{327} \left(c_{{\bm k},x,t,\sigma}^\dagger c_{{\bm k},x,b,\sigma} + \mathrm{H.c.}\right) \\
        +& \sum_{{\bm k}, \sigma} t_{\perp z}^{327} \left(c_{{\bm k},z,t,\sigma}^\dagger c_{{\bm k},z,b,\sigma} + \mathrm{H.c.}\right).
    \end{aligned}
\end{equation}
Here, \(c_{{\bm k},a,l,\sigma}^\dagger\) (\(c_{{\bm k},a,l,\sigma}\)) creates (annihilates) an electron with momentum \({\bm k}\), orbital \(a=x,z\), layer index \(l=t,b\), and spin \(\sigma\), where \(t\) and \(b\) denote the top and bottom Ni layers of the bilayer \ce{La3Ni2O7} block, respectively. The corresponding tight-binding parameters are listed in Table~\ref{tab:327_hopping_parameters}.

The local interaction term of the \ce{La3Ni2O7} block is taken in the same multiorbital Hubbard-Hund form as that used for the \ce{La2NiO4} layer,
\begin{equation}
    \label{eq:appendix_327_H_03}
    \begin{aligned}
        H_{\mathrm{int}}^{327} =&\ U^{327} \sum_{i,l} \left( n_{i,x,l,\uparrow} n_{i,x,l,\downarrow} + n_{i,z,l,\uparrow} n_{i,z,l,\downarrow} \right) \\
        &+ U^{\prime\,327} \sum_{i,l} n_{i,x,l} n_{i,z,l} \\
        &- J_H^{327} \sum_{i,l} \left( n_{i,x,l,\uparrow} n_{i,z,l,\uparrow} + n_{i,x,l,\downarrow} n_{i,z,l,\downarrow} \right) \\
        &+ J_H^{327} \sum_{i,l,\sigma} c_{i,x,l,\sigma}^{\dagger} c_{i,z,l,\bar{\sigma}}^{\dagger} c_{i,x,l,\bar{\sigma}} c_{i,z,l,\sigma} \\
        &+ J_C^{327} \sum_{i,l} \left( c_{i,x,l,\uparrow}^{\dagger} c_{i,x,l,\downarrow}^{\dagger} c_{i,z,l,\downarrow} c_{i,z,l,\uparrow} + \mathrm{H.c.} \right),
    \end{aligned}
\end{equation}
where $n_{i,a,l,\sigma}=c_{i,a,l,\sigma}^{\dagger}c_{i,a,l,\sigma}$ and $n_{i,a,l}=n_{i,a,l,\uparrow}+n_{i,a,l,\downarrow}$. Here, $i$ labels the in-plane lattice sites, \(a=x,z\) denotes the orbital index, and \(l=t,b\) labels the top and bottom Ni layers of the bilayer \ce{La3Ni2O7} block. The interaction parameters satisfy the rotationally invariant relations
\begin{equation}
    \label{eq:appendix_327_H_04}
    U^{\prime\,327}=U^{327}-2J_H^{327}, \qquad J_C^{327}=J_H^{327}.
\end{equation}
In the numerical calculations, we fix $J_H^{327}/U^{327}=0.1$, consistent with the choice used for the \ce{La2NiO4} layer.

The interacting \ce{La3Ni2O7} model is solved within the RISB framework at the saddle-point level, in the same spirit as described in Appendix~\ref{Sec:A1}. In the present work, however, we restrict the \ce{La3Ni2O7} sector to the paramagnetic solution. This treatment is sufficient for our purpose, since the coupled-layer calculation in the main text is not intended to establish the full correlated phase diagram of \ce{La3Ni2O7} itself, but rather to determine whether the intermediate \ce{La2NiO4} layer can recover an appreciable low-energy contribution once hybridized with the \ce{La3Ni2O7} block. In this context, the \ce{La3Ni2O7} block serves as the reference low-energy metallic component of \ce{La5Ni3O11}.

In the coupled-layer calculations presented in the main text, we set \(U^{327}=4.0\) eV, the same value as that adopted for the \ce{La2NiO4} layer. This choice is made for consistency between the two sectors and allows for a direct comparison of their correlated effects within the same interaction scale.

\section{DFT details}
Our DFT calculations employ the Vienna ab-initio simulation package (VASP) code \cite{kresse1996efficient} with the projector augmented wave (PAW) method \cite{kresse1999ultrasoft}. In this work, we consider two different exchange-correlation functionals. One is the generalized gradient approximation (GGA) exchange-correlation functional and its Perdew-Burke-Ernzerhof (PBE) version \cite{perdew1996generalized}, the other is the widely used HSE06 hybrid functionals \cite{krukau2006influence} that combine the Hartree-Fock (HF) and Kohn-Sham (KS) theories \cite{becke1993density}. The cutoff energy for expanding the wave functions into a plane-wave basis is set to be 500 eV. The energy convergence criterion is 10$^{−8}$ eV. The $\Gamma$-centered 13$\times$13$\times$3 and 7$\times$7$\times$2 k-meshes are used in PBE and HSE06 calculations, respectively. Focusing on the partially occupied $e_g$ orbitals from the DFT calculations,  we use the Wannier90 code \cite{mostofi2008wannier90,marzari2012maximally} to fit the $e_g$ bands and extract the tight-binding (TB) model parameters.
%references
%\bibliographystyle{plainnat} % 设置参考文献样式为plainnat
%\bibliography{references}
%\bibliographystyle{aps}

\bibliography{ref}

@Article{5410_magnetic_structure_01,
  author    = {LaBollita, Harrison and Botana, Antia S.},
  journal   = {Phys. Rev. B},
  title     = {Electronic structure and magnetic properties of higher-order layered nickelates: {${\mathrm{La}}_{n+1}{\mathrm{Ni}}_{n}{\mathrm{O}}_{2n+2}(n=4-6)$}},
  year      = {2021},
  month     = {Jul},
  pages     = {035148},
  volume    = {104},
  doi       = {10.1103/PhysRevB.104.035148},
  issue     = {3},
  numpages  = {11},
  publisher = {American Physical Society},
  url       = {https://link.aps.org/doi/10.1103/PhysRevB.104.035148},
}

@Article{327_original_exp_03,
  author   = {Li, Jingyuan and Peng, Di and Ma, Peiyue and Zhang, Hengyuan and Xing, Zhenfang and Huang, Xing and Huang, Chaoxin and Huo, Mengwu and Hu, Deyuan and Dong, Zixian and Chen, Xiang and Xie, Tao and Dong, Hongliang and Sun, Hualei and Zeng, Qiaoshi and Mao, Ho-kwang and Wang, Meng},
  journal  = {National Science Review},
  title    = {Identification of superconductivity in bilayer nickelate {$\mathrm{La}_{3}\mathrm{Ni}_{2}\mathrm{O}_{7}$} under high pressure up to 100 {GPa}},
  year     = {2025},
  issn     = {2095-5138},
  month    = {05},
  number   = {10},
  pages    = {nwaf220},
  volume   = {12},
  doi      = {10.1093/nsr/nwaf220},
  url      = {https://doi.org/10.1093/nsr/nwaf220},
}

@article{327_DFT_01,
  title = {Bilayer Two-Orbital Model of {$\mathrm{L}{\mathrm{a}}_{3}\mathrm{N}{\mathrm{i}}_{2}{\mathrm{O}}_{7}$} under Pressure},
  author = {Luo, Zhihui and Hu, Xunwu and Wang, Meng and W\'u, W\'ei and Yao, Dao-Xin},
  journal = {Phys. Rev. Lett.},
  volume = {131},
  issue = {12},
  pages = {126001},
  numpages = {6},
  year = {2023},
  month = {Sep},
  publisher = {American Physical Society},
  doi = {10.1103/PhysRevLett.131.126001},
  url = {https://link.aps.org/doi/10.1103/PhysRevLett.131.126001}
}

@article{327_DFT_02,
  title = {Electronic structure, dimer physics, orbital-selective behavior, and magnetic tendencies in the bilayer nickelate superconductor {$\mathrm{La}_{3}\mathrm{Ni}_{2}\mathrm{O}_{7}$} under pressure},
  author = {Zhang, Yang and Lin, Ling-Fang and Moreo, Adriana and Dagotto, Elbio},
  journal = {Phys. Rev. B},
  volume = {108},
  issue = {18},
  pages = {L180510},
  numpages = {5},
  year = {2023},
  month = {Nov},
  publisher = {American Physical Society},
  doi = {10.1103/PhysRevB.108.L180510},
  url = {https://link.aps.org/doi/10.1103/PhysRevB.108.L180510}
}

@Article{327_original_exp_01,
  author   = {Sun, Hualei and Huo, Mengwu and Hu, Xunwu and Li, Jingyuan and Liu, Zengjia and Han, Yifeng and Tang, Lingyun and Mao, Zhongquan and Yang, Pengtao and Wang, Bosen and Cheng, Jinguang and Yao, Dao-Xin and Zhang, Guang-Ming and Wang, Meng},
  journal  = {Nature},
  title    = {Signatures of superconductivity near 80{K} in a nickelate under high pressure},
  year     = {2023},
  issn     = {1476-4687},
  month    = {Sep},
  number   = {7979},
  pages    = {493-498},
  volume   = {621},
  day      = {01},
  doi      = {10.1038/s41586-023-06408-7},
  url      = {https://doi.org/10.1038/s41586-023-06408-7},
}

@Article{327_original_exp_02,
  author   = {Zhang, Yanan and Su, Dajun and Huang, Yanen and Shan, Zhaoyang and Sun, Hualei and Huo, Mengwu and Ye, Kaixin and Zhang, Jiawen and Yang, Zihan and Xu, Yongkang and Su, Yi and Li, Rui and Smidman, Michael and Wang, Meng and Jiao, Lin and Yuan, Huiqiu},
  journal  = {Nature Physics},
  title    = {High-temperature superconductivity with zero resistance and strange-metal behaviour in {$\mathrm{La}_{3}\mathrm{Ni}_{2}\mathrm{O}_{7-\delta}$}},
  year     = {2024},
  issn     = {1745-2481},
  month    = {Aug},
  number   = {8},
  pages    = {1269-1273},
  volume   = {20},
  day      = {01},
  doi      = {10.1038/s41567-024-02515-y},
  url      = {https://doi.org/10.1038/s41567-024-02515-y},
}

@Article{327_review_03,
  author   = {Wang, Yuxin and Jiang, Kun and Ying, Jianjun and Wu, Tao and Cheng, Jinguang and Hu, Jiangping and Chen, Xianhui},
  journal  = {National Science Review},
  title    = {Recent progress in nickelate superconductors},
  year     = {2025},
  issn     = {2095-5138},
  month    = {10},
  number   = {10},
  pages    = {nwaf373},
  volume   = {12},
  doi      = {10.1093/nsr/nwaf373},
  url      = {https://doi.org/10.1093/nsr/nwaf373},
}

@article{5311_structure_01,
  title = {Design and synthesis of three-dimensional hybrid {R}uddlesden-{P}opper nickelate single crystals},
  author = {Li, Feiyu and Guo, Ning and Zheng, Qiang and Shen, Yang and Wang, Shilei and Cui, Qihui and Liu, Chao and Wang, Shanpeng and Tao, Xutang and Zhang, Guang-Ming and Zhang, Junjie},
  journal = {Phys. Rev. Mater.},
  volume = {8},
  issue = {5},
  pages = {053401},
  numpages = {9},
  year = {2024},
  month = {May},
  publisher = {American Physical Society},
  doi = {10.1103/PhysRevMaterials.8.053401},
  url = {https://link.aps.org/doi/10.1103/PhysRevMaterials.8.053401}
}

@Article{5311_sc_01,
author={Shi, Mengzhu
and Peng, Di
and Fan, Kaibao
and Xing, Zhenfang
and Yang, Shaohua
and Wang, Yuzhu
and Li, Houpu
and Wu, Rongqi
and Du, Mei
and Ge, Binghui
and Zeng, Zhidan
and Zeng, Qiaoshi
and Ying, Jianjun
and Wu, Tao
and Chen, Xianhui},
title={Pressure induced superconductivity in hybrid {R}uddlesden‒{P}opper {$\mathrm{La}_{5}\mathrm{Ni}_{3}\mathrm{O}_{11}$} single crystals},
journal={Nature Physics},
year={2025},
month={Nov},
day={01},
volume={21},
number={11},
pages={1780-1786},
issn={1745-2481},
doi={10.1038/s41567-025-03023-3},
url={https://doi.org/10.1038/s41567-025-03023-3}
}

@article{5311_theory_RPA_01,
  title = {Electronic structure, and magnetic and superconducting pairing tendencies of the alternating single layer--bilayer stacking nickelate {$\mathrm{La}_{5}\mathrm{Ni}_{3}\mathrm{O}_{11}$} under pressure},
  author = {Zhang, Yang and Lin, Ling-Fang and Moreo, Adriana and Okamoto, Satoshi and Maier, Thomas A. and Dagotto, Elbio},
  journal = {Phys. Rev. B},
  volume = {112},
  issue = {9},
  pages = {094515},
  numpages = {11},
  year = {2025},
  month = {Sep},
  publisher = {American Physical Society},
  doi = {10.1103/1mr2-s6s8},
  url = {https://link.aps.org/doi/10.1103/1mr2-s6s8}
}

@Article{5311_theory_RPA_02,
author={Zhang, Ming
and Chen, Cui-Qun
and Yao, Dao-Xin
and Yang, Fan},
title={Pairing mechanism and superconductivity in pressurized {$\mathrm{La}_{5}\mathrm{Ni}_{3}\mathrm{O}_{11}$}},
journal={Science China Physics, Mechanics {\&} Astronomy},
year={2026},
month={Feb},
day={06},
volume={69},
number={5},
pages={257411},
issn={1869-1927},
doi={10.1007/s11433-025-2907-0},
url={https://doi.org/10.1007/s11433-025-2907-0}
}

@misc{5311_theory_DMFT_01,
      title={Correlated electronic structure of the alternating monolayer-bilayer nickelate {$\mathrm{La}_{5}\mathrm{Ni}_{3}\mathrm{O}_{11}$}}, 
      author={Harrison LaBollita and Antia S. Botana},
      year={2026},
      eprint={2505.07394},
      archivePrefix={arXiv},
      primaryClass={cond-mat.str-el},
      url={https://arxiv.org/abs/2505.07394}, 
}

@misc{5311_theory_DMFT_02,
      title={Electronic structure, quasiparticle renormalizations, and magnetic correlations in the alternating single-layer bilayer nickelate {$\mathrm{La}_{5}\mathrm{Ni}_{3}\mathrm{O}_{11}$}}, 
      author={I. V. Leonov},
      year={2026},
      eprint={2604.26627},
      archivePrefix={arXiv},
      primaryClass={cond-mat.str-el},
      url={https://arxiv.org/abs/2604.26627}, 
}

@article{5311_theory_DMFT_03,
  title = {Phase diagrams and two key factors to superconductivity of {R}uddlesden-{P}opper nickelates},
  author = {Ouyang, Zhenfeng and He, Rong-Qiang and Lu, Zhong-Yi},
  journal = {Phys. Rev. B},
  volume = {112},
  issue = {4},
  pages = {045127},
  numpages = {10},
  year = {2025},
  month = {Jul},
  publisher = {American Physical Society},
  doi = {10.1103/1412-nfzm},
  url = {https://link.aps.org/doi/10.1103/1412-nfzm}
}

@article{327_phase_diagram_01,
  title = {Pressure-Induced Superconductivity In Polycrystalline {$\mathrm{La}_{3}\mathrm{Ni}_{2}\mathrm{O}_{7-\delta}$}},
  author = {Wang, G. and Wang, N. N. and Shen, X. L. and Hou, J. and Ma, L. and Shi, L. F. and Ren, Z. A. and Gu, Y. D. and Ma, H. M. and Yang, P. T. and Liu, Z. Y. and Guo, H. Z. and Sun, J. P. and Zhang, G. M. and Calder, S. and Yan, J.-Q. and Wang, B. S. and Uwatoko, Y. and Cheng, J.-G.},
  journal = {Phys. Rev. X},
  volume = {14},
  issue = {1},
  pages = {011040},
  numpages = {8},
  year = {2024},
  month = {Mar},
  publisher = {American Physical Society},
  doi = {10.1103/PhysRevX.14.011040},
  url = {https://link.aps.org/doi/10.1103/PhysRevX.14.011040}
}

@article{327_FLEX_01,
  title = {Possible High ${T}_{c}$ Superconductivity in {$\mathrm{La}_{3}\mathrm{Ni}_{2}\mathrm{O}_{7}$} under High Pressure through Manifestation of a Nearly Half-Filled Bilayer Hubbard Model},
  author = {Sakakibara, Hirofumi and Kitamine, Naoya and Ochi, Masayuki and Kuroki, Kazuhiko},
  journal = {Phys. Rev. Lett.},
  volume = {132},
  issue = {10},
  pages = {106002},
  numpages = {6},
  year = {2024},
  month = {Mar},
  publisher = {American Physical Society},
  doi = {10.1103/PhysRevLett.132.106002},
  url = {https://link.aps.org/doi/10.1103/PhysRevLett.132.106002}
}

@Article{4310_original_exp_01,
author={Zhu, Yinghao
and Peng, Di
and Zhang, Enkang
and Pan, Bingying
and Chen, Xu
and Chen, Lixing
and Ren, Huifen
and Liu, Feiyang
and Hao, Yiqing
and Li, Nana
and Xing, Zhenfang
and Lan, Fujun
and Han, Jiyuan
and Wang, Junjie
and Jia, Donghan
and Wo, Hongliang
and Gu, Yiqing
and Gu, Yimeng
and Ji, Li
and Wang, Wenbin
and Gou, Huiyang
and Shen, Yao
and Ying, Tianping
and Chen, Xiaolong
and Yang, Wenge
and Cao, Huibo
and Zheng, Changlin
and Zeng, Qiaoshi
and Guo, Jian-gang
and Zhao, Jun},
title={Superconductivity in pressurized trilayer {$\mathrm{La}_{4}\mathrm{Ni}_{3}\mathrm{O}_{10-\delta}$} single crystals},
journal={Nature},
year={2024},
month={Jul},
day={01},
volume={631},
number={8021},
pages={531-536},
issn={1476-4687},
doi={10.1038/s41586-024-07553-3},
url={https://doi.org/10.1038/s41586-024-07553-3}
}

@article{4310_original_exp_02,
  title = {Superconductivity in Trilayer Nickelate {${\mathrm{La}}_{4}{\mathrm{Ni}}_{3}{\mathrm{O}}_{10}$} under Pressure},
  author = {Zhang, Mingxin and Pei, Cuiying and Peng, Di and Du, Xian and Hu, Weixiong and Cao, Yantao and Wang, Qi and Wu, Juefei and Li, Yidian and Liu, Huanyu and Wen, Chenhaoping and Song, Jing and Zhao, Yi and Li, Changhua and Cao, Weizheng and Zhu, Shihao and Zhang, Qing and Yu, Na and Cheng, Peihong and Zhang, Lili and Li, Zhiwei and Zhao, Jinkui and Chen, Yulin and Jin, Changqing and Guo, Hanjie and Wu, Congjun and Yang, Fan and Zeng, Qiaoshi and Yan, Shichao and Yang, Lexian and Qi, Yanpeng},
  journal = {Phys. Rev. X},
  volume = {15},
  issue = {2},
  pages = {021005},
  numpages = {11},
  year = {2025},
  month = {Apr},
  publisher = {American Physical Society},
  doi = {10.1103/PhysRevX.15.021005},
  url = {https://link.aps.org/doi/10.1103/PhysRevX.15.021005}
}

@article{4310_original_exp_03,
doi = {10.1088/0256-307X/41/1/017401},
url = {https://doi.org/10.1088/0256-307X/41/1/017401},
year = {2024},
month = {jan},
publisher = {Chinese Physical Society and IOP Publishing Ltd},
volume = {41},
number = {1},
pages = {017401},
author = {Li, Qing and Zhang, Ying-Jie and Xiang, Zhe-Ning and Zhang, Yuhang and Zhu, Xiyu and Wen, Hai-Hu},
title = {Signature of Superconductivity in Pressurized {$\mathrm{La}_{4}\mathrm{Ni}_{3}\mathrm{O}_{10}$}},
journal = {Chinese Physics Letters},
}

@article{214_magnetic_01,
doi = {10.1088/0953-8984/3/19/002},
url = {https://doi.org/10.1088/0953-8984/3/19/002},
year = {1991},
month = {may},
publisher = {},
volume = {3},
number = {19},
pages = {3215},
author = {J Rodriguez-Carvajal and M T Fernandez-Diaz and J L Martinez},
title = {Neutron diffraction study on structural and magnetic properties of {$\mathrm{La}_{2}\mathrm{Ni}\mathrm{O}_{4}$}},
journal = {Journal of Physics: Condensed Matter},
}

@article{214_photoemission_01,
  title = {Electronic structure of {$\mathrm{La}_{2-x}\mathrm{Sr}_{x}\mathrm{NiO}_{4}$} studied by photoemission and inverse-photoemission spectroscopy},
  author = {Eisaki, H. and Uchida, S. and Mizokawa, T. and Namatame, H. and Fujimori, A. and Sasagawa, T. and Kishio, K. and Kitazawa, K. and Kojima, H. and Tanaka, S.},
  journal = {Phys. Rev. B},
  volume = {45},
  issue = {21},
  pages = {12513--12521},
  year = {1992},
  month = {Jun},
  publisher = {American Physical Society},
  doi = {10.1103/PhysRevB.45.12513},
  url = {https://doi.org/10.1103/PhysRevB.45.12513},
}

@article{214_magnetic_02,
  title = {Magnetic Correlations in {${\mathrm{La}}_{2}\mathrm{Ni}{\mathrm{O}}_{4+\ensuremath{\delta}}$}},
  author = {Aeppli, G. and Buttrey, D. J.},
  journal = {Phys. Rev. Lett.},
  volume = {61},
  issue = {2},
  pages = {203--206},
  numpages = {0},
  year = {1988},
  month = {Jul},
  publisher = {American Physical Society},
  doi = {10.1103/PhysRevLett.61.203},
  url = {https://link.aps.org/doi/10.1103/PhysRevLett.61.203}
}

@article{214_magnetic_03,
title = {Magnetic structure and weak ferromagnetism of {${\mathrm{La}}_{2}\mathrm{Ni}{\mathrm{O}}_{4+\ensuremath{\delta}}$}},
journal = {Physica C: Superconductivity},
volume = {191},
number = {1},
pages = {15-22},
year = {1992},
issn = {0921-4534},
doi = {https://doi.org/10.1016/0921-4534(92)90625-M},
url = {https://www.sciencedirect.com/science/article/pii/092145349290625M},
author = {K. Yamada and T. Omata and K. Nakajima and S. Hosoya and T. Sumida and Y. Endoh},
}

@article{214_magnetic_04,
    author = {Wang, X. L. and Stassis, C. and Johnston, D. C. and Leung, T. C. and Ye, J. and Harmon, B. N. and Lander, G. H. and Schultz, A. J. and    Loong, C.-K. and Honig, J. M.},
    title = {The antiferromagnetic form factor of {$\mathrm{La}_{2}\mathrm{Ni}\mathrm{O}_{4}$}},
    journal = {Journal of Applied Physics},
    volume = {69},
    number = {8},
    pages = {4860-4862},
    year = {1991},
    month = {04},
    issn = {0021-8979},
    doi = {10.1063/1.348204},
    url = {https://doi.org/10.1063/1.348204},
}

@Article{214_magnetic_05,
  title = {High-energy spin waves in the spin-1 square-lattice antiferromagnet {$\mathrm{La}_{2}\mathrm{Ni}\mathrm{O}_{4}$}},
  author = {Petsch, A. N. and Headings, N. S. and Prabhakaran, D. and Kolesnikov, A. I. and Frost, C. D. and Boothroyd, A. T. and Coldea, R. and Hayden, S. M.},
  journal = {Phys. Rev. Res.},
  volume = {5},
  issue = {3},
  pages = {033113},
  numpages = {11},
  year = {2023},
  month = {Aug},
  publisher = {American Physical Society},
  doi = {10.1103/PhysRevResearch.5.033113},
  url = {https://link.aps.org/doi/10.1103/PhysRevResearch.5.033113}
}

@misc{214_magnetic_06,
      title={Pressure and doping control of magnetic order and metallization in {R}uddlesden-{P}opper {$\mathrm{La}_{2}\mathrm{Ni}\mathrm{O}_{4}$}}, 
      author={Han-Yu Wang and Shu-Hong Tang and Xiao-Teng Huang and Ya-Min Quan and XianLong Wang and Yan-Ling Li and Da-Yong Liu and H. -Q. Lin and Zhi Zeng and Liang-Jian Zou},
      year={2026},
      eprint={2507.03277},
      archivePrefix={arXiv},
      primaryClass={cond-mat.supr-con},
      url={https://arxiv.org/abs/2507.03277}, 
}

@article{KRSB_01,
  title = {New Functional Integral Approach to Strongly Correlated Fermi Systems: The Gutzwiller Approximation as a Saddle Point},
  author = {Kotliar, Gabriel and Ruckenstein, Andrei E.},
  journal = {Phys. Rev. Lett.},
  volume = {57},
  issue = {11},
  pages = {1362--1365},
  numpages = {0},
  year = {1986},
  month = {Sep},
  publisher = {American Physical Society},
  doi = {10.1103/PhysRevLett.57.1362},
  url = {https://link.aps.org/doi/10.1103/PhysRevLett.57.1362}
}

@article{RISB_01,
  title = {Rotationally invariant slave-boson formalism and momentum dependence of the quasiparticle weight},
  author = {Lechermann, Frank and Georges, Antoine and Kotliar, Gabriel and Parcollet, Olivier},
  journal = {Phys. Rev. B},
  volume = {76},
  issue = {15},
  pages = {155102},
  numpages = {20},
  year = {2007},
  month = {Oct},
  publisher = {American Physical Society},
  doi = {10.1103/PhysRevB.76.155102},
  url = {https://link.aps.org/doi/10.1103/PhysRevB.76.155102}
}

@article{RISB_02,
  title = {Rotationally invariant slave bosons for strongly correlated superconductors},
  author = {Isidori, A. and Capone, M.},
  journal = {Phys. Rev. B},
  volume = {80},
  issue = {11},
  pages = {115120},
  numpages = {18},
  year = {2009},
  month = {Sep},
  publisher = {American Physical Society},
  doi = {10.1103/PhysRevB.80.115120},
  url = {https://link.aps.org/doi/10.1103/PhysRevB.80.115120}
}

@Article{Hund_coupling_01,
author={W{\'u}, W{\'e}i
and Luo, Zhihui
and Yao, Dao-Xin
and Wang, Meng},
title={Superexchange and charge transfer in the nickelate superconductor {$\mathrm{La}_{3}\mathrm{Ni}_{2}\mathrm{O}_{7}$} under pressure},
journal={Science China Physics, Mechanics {\&} Astronomy},
year={2024},
month={Mar},
day={26},
volume={67},
number={11},
pages={117402},
issn={1869-1927},
doi={10.1007/s11433-023-2300-4},
url={https://doi.org/10.1007/s11433-023-2300-4}
}

@Article{327_ARPES_01,
author={Yang, Jiangang
and Sun, Hualei
and Hu, Xunwu
and Xie, Yuyang
and Miao, Taimin
and Luo, Hailan
and Chen, Hao
and Liang, Bo
and Zhu, Wenpei
and Qu, Gexing
and Chen, Cui-Qun
and Huo, Mengwu
and Huang, Yaobo
and Zhang, Shenjin
and Zhang, Fengfeng
and Yang, Feng
and Wang, Zhimin
and Peng, Qinjun
and Mao, Hanqing
and Liu, Guodong
and Xu, Zuyan
and Qian, Tian
and Yao, Dao-Xin
and Wang, Meng
and Zhao, Lin
and Zhou, X. J.},
title={Orbital-dependent electron correlation in double-layer nickelate {$\mathrm{La}_{3}\mathrm{Ni}_{2}\mathrm{O}_{7}$}},
journal={Nature Communications},
year={2024},
month={May},
day={23},
volume={15},
number={1},
pages={4373},
issn={2041-1723},
doi={10.1038/s41467-024-48701-7},
url={https://doi.org/10.1038/s41467-024-48701-7}
}

@article{327_ARPES_02,
doi = {10.1088/0256-307X/41/8/087402},
url = {https://doi.org/10.1088/0256-307X/41/8/087402},
year = {2024},
month = {jul},
publisher = {Chinese Physical Society and IOP Publishing Ltd},
volume = {41},
number = {8},
pages = {087402},
author = {Li, Yidian and Du, Xian and Cao, Yantao and Pei, Cuiying and Zhang, Mingxin and Zhao, Wenxuan and Zhai, Kaiyi and Xu, Runzhe and Liu, Zhongkai and Li, Zhiwei and Zhao, Jinkui and Li, Gang and Qi, Yanpeng and Guo, Hanjie and Chen, Yulin and Yang, Lexian},
title = {Electronic Correlation and Pseudogap-Like Behavior of High-Temperature Superconductor {$\mathrm{La}_{3}\mathrm{Ni}_{2}\mathrm{O}_{7}$}},
journal = {Chinese Physics Letters},
}

@article{327_DMFT_01,
  title = {Quenched Pair Breaking by Interlayer Correlations as a Key to Superconductivity in {$\mathrm{La}_{3}\mathrm{Ni}_{2}\mathrm{O}_{7}$}},
  author = {Ryee, Siheon and Witt, Niklas and Wehling, Tim O.},
  journal = {Phys. Rev. Lett.},
  volume = {133},
  issue = {9},
  pages = {096002},
  numpages = {7},
  year = {2024},
  month = {Aug},
  publisher = {American Physical Society},
  doi = {10.1103/PhysRevLett.133.096002},
  url = {https://link.aps.org/doi/10.1103/PhysRevLett.133.096002}
}

@Article{327_Pr_01,
author={Wang, Ningning
and Wang, Gang
and Shen, Xiaoling
and Hou, Jun
and Luo, Jun
and Ma, Xiaoping
and Yang, Huaixin
and Shi, Lifen
and Dou, Jie
and Feng, Jie
and Yang, Jie
and Shi, Yunqing
and Ren, Zhian
and Ma, Hanming
and Yang, Pengtao
and Liu, Ziyi
and Liu, Yue
and Zhang, Hua
and Dong, Xiaoli
and Wang, Yuxin
and Jiang, Kun
and Hu, Jiangping
and Nagasaki, Shoko
and Kitagawa, Kentaro
and Calder, Stuart
and Yan, Jiaqiang
and Sun, Jianping
and Wang, Bosen
and Zhou, Rui
and Uwatoko, Yoshiya
and Cheng, Jinguang},
title={Bulk high-temperature superconductivity in pressurized tetragonal {$\mathrm{La}_{2}\mathrm{PrNi}_{2}\mathrm{O}_{7}$}},
journal={Nature},
year={2024},
month={Oct},
day={01},
volume={634},
number={8034},
pages={579-584},
issn={1476-4687},
doi={10.1038/s41586-024-07996-8},
url={https://doi.org/10.1038/s41586-024-07996-8}
}

@Article{327_Sm_01,
author={Li, Feiyu
and Xing, Zhenfang
and Peng, Di
and Dou, Jie
and Guo, Ning
and Ma, Liang
and Zhang, Yulin
and Wang, Lingzhen
and Luo, Jun
and Yang, Jie
and Zhang, Jian
and Chang, Tieyan
and Chen, Yu-Sheng
and Cai, Weizhao
and Cheng, Jinguang
and Wang, Yuzhu
and Liu, Yuxin
and Luo, Tao
and Hirao, Naohisa
and Matsuoka, Takahiro
and Kadobayashi, Hirokazu
and Zeng, Zhidan
and Zheng, Qiang
and Zhou, Rui
and Zeng, Qiaoshi
and Tao, Xutang
and Zhang, Junjie},
title={Bulk superconductivity up to 96 {K} in pressurized nickelate single crystals},
journal={Nature},
year={2026},
month={Jan},
day={01},
volume={649},
number={8098},
pages={871-878},
issn={1476-4687},
doi={10.1038/s41586-025-09954-4},
url={https://doi.org/10.1038/s41586-025-09954-4}
}

@article{FeSe_intercalation_01,
  title = {Enhancement of the superconducting transition temperature of FeSe by intercalation of a molecular spacer layer},
  author = {Burrard-Lucas, M. and Free, D. G. and Sedlmaier, S. J. and Wright, J. D. and Cassidy, S. J. and Hara, Y. and Corkett, A. J. and Lancaster, T. and Baker, P. J. and Blundell, S. J. and Clarke, S. J.},
  journal = {Nature Materials},
  volume = {12},
  pages = {15--19},
  year = {2013},
  doi = {10.1038/nmat3464},
  url = {https://doi.org/10.1038/nmat3464},
}

@article{FeSe_intercalation_02,
  title = {Superconductivity at 30 K in the vicinity of the insulating phase in {${\mathrm{K}}_{x}{\mathrm{Fe}}_{2-y}{\mathrm{Se}}_{2}$}},
  author = {Guo, Jiangang and Jin, Sheng and Wang, Gang and Wang, Shifeng and Zhu, Kaixing and Zhou, Tingting and He, Meng and Chen, Xiaolong},
  journal = {Phys. Rev. B},
  volume = {82},
  issue = {18},
  pages = {180520},
  year = {2010},
  doi = {10.1103/PhysRevB.82.180520},
  url = {https://doi.org/10.1103/PhysRevB.82.180520},
}

@article{kresse1996efficient,
   title = {Efficient iterative schemes for ab initio total-energy calculations using a plane-wave basis set},
  author = {Kresse, G. and Furthm\"uller, J.},
  journal = {Phys. Rev. B},
  volume = {54},
  issue = {16},
  pages = {11169--11186},
  numpages = {0},
  year = {1996},
  month = {Oct},
  publisher = {American Physical Society},
  doi = {10.1103/PhysRevB.54.11169},
  url = {https://link.aps.org/doi/10.1103/PhysRevB.54.11169}
}

@article{kresse1999ultrasoft,
  title = {From ultrasoft pseudopotentials to the projector augmented-wave method},
  author = {Kresse, G. and Joubert, D.},
  journal = {Phys. Rev. B},
  volume = {59},
  issue = {3},
  pages = {1758--1775},
  numpages = {0},
  year = {1999},
  month = {Jan},
  publisher = {American Physical Society},
  doi = {10.1103/PhysRevB.59.1758},
  url = {https://link.aps.org/doi/10.1103/PhysRevB.59.1758}
}

@article{perdew1996generalized,
  title = {Generalized Gradient Approximation Made Simple},
  author = {Perdew, John P. and Burke, Kieron and Ernzerhof, Matthias},
  journal = {Phys. Rev. Lett.},
  volume = {77},
  issue = {18},
  pages = {3865--3868},
  numpages = {0},
  year = {1996},
  month = {Oct},
  publisher = {American Physical Society},
  doi = {10.1103/PhysRevLett.77.3865},
  url = {https://link.aps.org/doi/10.1103/PhysRevLett.77.3865}
}

@article{becke1993density,
  title="{Density‐functional thermochemistry. III. The role of exact exchange}",
  author={Becke, Axel D},
  journal={The Journal of chemical physics},
  volume={98},
  number={7},
  pages={5648-5652},
  year={1993},
  publisher={American Institute of Physics},
}

@article{krukau2006influence,
  title={Influence of the exchange screening parameter on the performance of screened hybrid functionals},
  author={Krukau, Aliaksandr V and Vydrov, Oleg A and Izmaylov, Artur F and Scuseria, Gustavo E},
  journal={The Journal of chemical physics},
  volume={125},
  number={22},
  year={2006},
  publisher={AIP Publishing},
  url={https://pubs.aip.org/aip/jcp/article/125/22/224106/953719}
}

@article{mostofi2008wannier90,
  title="{wannier90: A tool for obtaining maximally-localised Wannier functions}",
  author={Mostofi, Arash A and Yates, Jonathan R and Lee, Young-Su and Souza, Ivo and Vanderbilt, David and Marzari, Nicola},
  journal={Computer physics communications},
  volume={178},
  number={9},
  pages={685--699},
  year={2008},
  publisher={Elsevier},
  url={https://www.sciencedirect.com/science/article/pii/S0010465507004936}
}

@article{marzari2012maximally,
  title = {Maximally localized Wannier functions: Theory and applications},
  author = {Marzari, Nicola and Mostofi, Arash A. and Yates, Jonathan R. and Souza, Ivo and Vanderbilt, David},
  journal = {Rev. Mod. Phys.},
  volume = {84},
  issue = {4},
  pages = {1419--1475},
  numpages = {0},
  year = {2012},
  month = {Oct},
  publisher = {American Physical Society},
  doi = {10.1103/RevModPhys.84.1419},
  url = {https://link.aps.org/doi/10.1103/RevModPhys.84.1419}
}

@article{takegahara1989electronic,
  title="{Electronic band structure of paramagnetic and antiferromagnetic La$_2$NiO$_4$}",
  author={Takegahara, Katsuhiko and Kasuya, Tadao},
  journal={Solid state communications},
  volume={70},
  number={6},
  pages={641--645},
  year={1989},
  publisher={Elsevier},
  url={https://www.sciencedirect.com/science/article/abs/pii/0038109889903657}
}

@article{rivero2009description,
  title={Description of magnetic interactions in strongly correlated solids via range-separated hybrid functionals},
  author={Rivero, Pablo and Moreira, Iberio de PR and Scuseria, Gustavo E and Illas, Francesc},
  journal={Physical Review B—Condensed Matter and Materials Physics},
  volume={79},
  number={24},
  pages={245129},
  year={2009},
  publisher={APS},
  url={https://journals.aps.org/prb/abstract/10.1103/PhysRevB.79.245129}
}

@Article{327_ARPES_03,
author={Nie, Zihao
and Li, Yueying
and Lv, Wei
and Xu, Lizhi
and Jiang, Zhicheng
and Fu, Peng
and Zhou, Guangdi
and Song, Wenhua
and Chen, Yaqi
and Wang, Heng
and Huang, Haoliang
and Lin, Junhao
and Jia, Jin-Feng
and Shen, Dawei
and Li, Peng
and Xue, Qi-Kun
and Chen, Zhuoyu},
title={Superconductivity and electronic structures of nickelate thin film superstructures},
journal={Nature},
year={2026},
month={Apr},
day={01},
volume={652},
number={8110},
pages={628-634},
issn={1476-4687},
doi={10.1038/s41586-026-10352-7},
url={https://doi.org/10.1038/s41586-026-10352-7}
}

\end{document}